\documentclass[epj]{svjour}

\usepackage{graphics}

\usepackage{graphicx}
\usepackage{amsmath}
\usepackage{amssymb}
\usepackage{relsize}
\usepackage{upgreek}

\usepackage[numbers]{natbib}

\renewcommand{\vec}[1]{\boldsymbol{#1}}
\newcommand{\shearrate}{\dot\gamma}
\newcommand{\axis}{a}
\newcommand{\vol}{V}
\newcommand{\vel}{v}
\newcommand{\pos}{x}
\newcommand{\basis}{e}
\newcommand{\incl}{\theta}
\newcommand{\phase}{\phi}
\newcommand{\visc}{\eta}
\newcommand{\viscin}{\visc^\text{i}}
\newcommand{\viscout}{\visc^\text{o}}
\newcommand{\energy}{E}
\newcommand{\elast}{\energy_0}
\newcommand{\elastprop}{\epsilon_0}
\newcommand{\tim}{t}
\newcommand{\f}{f}

\newcommand{\energyscale}{\elast}
\newcommand{\dimvisc}{\lambda}

\newcommand{\dimshear}{\chi}
\newcommand{\diminvshear}{\chi^{-1}}
\newcommand{\ecc}{\alpha}
\newcommand{\dimtim}{\tau}
\newcommand{\su}{\Sigma}
\newcommand{\di}{\Delta}
\newcommand{\slope}{\rho}
\newcommand{\meanslope}{\bar{\rho}}
\newcommand{\amp}{b}
\newcommand{\freq}{\omega}
\newcommand{\dimfreq}{\tilde\freq}
\newcommand{\period}{T}
\newcommand{\ph}{\varphi}
\newcommand{\aph}{\nu}
\newcommand{\sm}{\varepsilon}

\newcommand{\tumb}{\langle\dot\incl\rangle}

\newcommand{\mean}[1]{\left\langle{#1}\right\rangle}
\newcommand{\abs}[1]{\left|{#1}\right|}
\newcommand{\ceil}[1]{\left\lceil{#1}\right\rceil}
\newcommand{\floor}[1]{\left\lfloor{#1}\right\rfloor}

\DeclareMathOperator{\erf}{erf}
\DeclareMathOperator{\erfi}{erfi}
\DeclareMathOperator{\erfc}{erfc}

\DeclareMathOperator{\sign}{sign}

\DeclareMathOperator{\re}{Re}
\DeclareMathOperator{\im}{Im}

\begin{document}

\title{Elastic capsules in shear flow:\\
Analytical solutions for constant and time-dependent shear rates}
\titlerunning{Elastic capsules in shear flow}

\author{Steffen Kessler \and Reimar Finken \and Udo Seifert}
\institute{II.\ Institut f{\"u}r Theoretische Physik,\\
  Pfaffenwaldring 57, \\Universit{\"a}t Stuttgart,\\ 70550 Stuttgart,\\
Germany}

\date{Draft of \today}
%
\abstract{We investigate the dynamics of microcapsules in linear shear flow
  within a reduced model with two degrees of freedom. In previous work for
  steady shear flow, the dynamic phases of this model, i.e. swinging,
  tumbling and intermittent behaviour, have been identified using numerical
  methods. In this paper, we integrate the equations of motion in the
  quasi-spherical limit analytically for time-constant and time-dependent
  shear flow using matched asymptotic expansions. Using this method, we find
  analytical expressions for the mean tumbling rate in general time-dependent
  shear flow. The capsule dynamics is studied in more detail when the inverse
  shear rate is harmonically modulated around a constant mean value for which
  a dynamic phase diagram is constructed. By a judicious choice of both
  modulation frequency and phase, tumbling motion can be induced even if the
  mean shear rate corresponds to the swinging regime. We derive expressions
  for the amplitude and width of the resonance peaks as a function of the
  modulation frequency.
  \PACS{ {87.16.D-}{Membranes, bilayers, and vesicles} \and
    {47.15.G-}{Low-Reynolds-number (creeping) flows} } 
} 
\maketitle
\section{Introduction}
\label{sec:introduction-1}

The dynamic motion of soft objects such as elastic microcapsules in shear flow
represents a long-standing problem in science and engineering. It has received
increasing interest recently, in particular due to its relevance to
biological, medicinal and microfluidic applications. This problem is
challenging from a theoretical point of view, because the shape of these
objects is not given \emph{a priori}, but determined dynamically from a
balance of interfacial forces with fluid stresses. New insight has been gained
due to a plethora of experimental \cite{eggleton1998, walter2001, fischer2004,
  abkarian2007, fischer2007}, theoretical \cite{barthes-biesel1980,
  barthes-biesel1981, keller1982, bart85, barthes-biesel2002, lac2004,
  lac2005, finken2006, skotheim2007}, and numerical \cite{pozrikidis1995,
  ramanujan1998, noguchi2005, kessler2007, sui2008, sui2008a, dodson2008}
methods.

Perhaps the most well-known dynamic state of initially spherical elastic
microcapsules in shear flow is the tank-treading motion also present in fluid
vesicles \cite{kraus1996, haas1997, seifert1999a, pozrikidis2001b,
  noguchi2004, kantsler2005a, kantsler2006a, misbah2006, vlahovska2007,
  lebedev2007}, as reviewed in the first two chapters of
\cite{pozrikidis2003book}. In contrast to fluid vesicles, microcapsules
exhibit a finite shear elasticity, since their membrane is chemically or
physically cross-linked.  This holds both for artificial polymerised capsules
\cite{walter2001} and red blood cells (RBCs), whose membrane is composed of an
incompressible lipid bilayer underlined by a thin elastic cytoskeleton
\cite{mohandas1994}. For a short time, viscous fluid vesicles can also resist
shear.

The resistance to shear leads to qualitatively different behaviour, such as
preventing the prolate to oblate shape transition of viscous fluid vesicles
\cite{noguchi2005}. Perhaps most surprisingly, it also leads to qualitatively
different instabilities like wrinkling first observed on polymerised capsules
\cite{walter2001,finken2006} and later as a transient on viscous vesicles
\cite{kantsler2007}.

When the unstressed initial shape of the cell is not spherical, material
elements of the membrane are deformed when displaced from their initial
position. This shape memory, suggested for RBCs in Ref.~\cite{fischer2004},
leads to an oscillation of the inclination angle superimposed on the
tank-treading motion, called swinging, and an intermittent regime between
tank-treading and tumbling \cite{abkarian2007,skotheim2007}. The swinging
motion of RBCs was studied numerically in Ref.~\cite{ramanujan1998} using a
boundary integral formulation of the hydrodynamics. Later, more comprehensive
studies of all dynamic phases were performed using both a spectral numerical
method \cite{kessler2007} and an immersed boundary lattice Boltzmann method
\cite{sui2008,sui2008a}. The phase diagram constructed in
Ref.~\cite{kessler2007} basically confirmed the qualitative correctness of a
reduced model \cite{skotheim2007} at low to moderate viscosity
ratios. However, both Refs.~\cite{kessler2007} and \cite{sui2008a}
independently contested the intermittent regime at large viscosity ratios as
an artifact of the reduced model. Instead, in these works the tumbling motion
was found to be a transient towards a stable swinging motion. Numerical
studies of elongated capsules in extensional flow at high flow rates reveal a
novel bifurcation between a spindled and a cusped capsule shape induced by
compressive stresses \cite{dodson2008}.

New phenomena are expected when the driving shear flow is no longer constant
in time and space. Indeed, a transient wrinkling phenomenon was observed for
fluid vesicles in suddenly reversed elongational shear flow
\cite{kantsler2007}, where the stress becomes momentarily compressive. For
spatially varying shear flow produced by structured microchannels, a
transition from prolate shape to bullet-like shape as well as symmetry
breaking transitions were observed in vesicles both experimentally and in
simulations \cite{noguchi2008}.

So far, microcapsule dynamics has only been studied in steady shear flow. It
is the aim of this paper to investigate the effects of modulating the shear
rate on the dynamics of capsules.  In particular, we want to focus on the
question whether dynamic phase transitions can be induced by small amplitude
oscillations around a fixed mean shear rate. Since solving the exact equations
of motion numerically is computationally prohibitive, we constrain our
investigations to the reduced model of Skotheim et al.~\cite{skotheim2007},
which will allow analytical solutions in the quasi-spherical limit.

This paper is structured as follows: After reviewing the reduced model
\cite{skotheim2007}, we derive non-dimensional equations of motion valid in
the quasi-spherical limit in section~\ref{sec:reduced-model}. A numerical
study of the dynamics for constant shear rate yields the dynamic phase
diagram. In the quasi-spherical limit the equations of motion can be
integrated exactly for time-constant shear flow, which is done in section
\ref{sec:constant-shear-rate}. Integration is even possible for general
time-dependent shear flow, which is studied in
section~\ref{sec:time-modul-shear} in detail. Specialising to harmonic
modulations of the shear rate around a constant mean value reveals that
tumbling motion can be induced from the swinging regime at certain resonance
frequencies of the modulation. Both the resonance behaviour and the smooth
off-resonance background are studied in detail. A dynamic phase diagram as a
function of modulation amplitude and frequency is constructed. The more
intricate details of the calculations are shown in the Appendices.

\section{Reduced model}
\label{sec:reduced-model}

\begin{figure}[t!]
  \centering
  \includegraphics[width=0.9\linewidth]{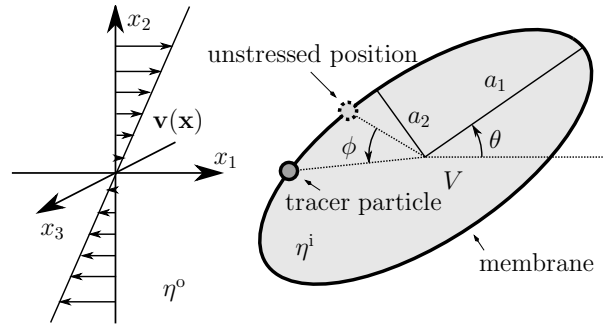}
  \caption{2d-cut of a 3d-ellipsoidal capsule with semi-axes $\axis_i$ and
    volume $\vol$ orientated in an external linear shear flow with outer
    viscosity $\viscout$ encapsulating a fluid with inner viscosity
    $\viscin$. The inclination angle $\incl$ measures the angle between the
    direction of the long axis and the direction of the shear flow. The phase
    angle $\phase$ measures the tank-treading motion.}
  \label{fig:capsule}
\end{figure}

\subsection{Equation of motion}
\label{sec:equation-of-motion}
We investigate a reduced model of an elastic capsule with fixed ellipsoidal
shape (semi-axes $\axis_1$, $\axis_2$, $\axis_3$, and volume $\vol$) in a
linear shear flow with velocity
\begin{equation}
  \vec \vel (\vec\pos) = \shearrate \pos_2 \vec \basis_1 \,.
\end{equation}
depending upon position $\vec\pos=\sum_i x_i\vec\basis_i$ and shear rate
$\shearrate$.  The two axes $\axis_1$ and $\axis_2$ lie in the shear plane
with $\axis_1\geq\axis_2$ (see Fig.~\ref{fig:capsule}). Thus the axis
$\axis_3$ is perpendicular to the shear plane. The long axis $\axis_1$ is
oriented with inclination angle $\incl$ with respect to the direction of the
shear flow. The inner and outer flow have viscosities $\viscin$ and
$\viscout$, respectively. The membrane can tank-tread with respect to the
fixed ellipsoidal shape, measured by the phase angle $\phase$. This is the
Keller-Skalak model for a Red Blood Cell \cite{keller1982}.
Abkarian et al.~\cite{abkarian2007} and Skotheim and Secomb
\cite{skotheim2007} add an elastic energy term ($\elast\sin^2\phase$) which is
due to the tank-treading motion and the shape memory effect
\cite{fischer2004}. Abkarian et al.~\cite{abkarian2007} also consider a
viscosity of the membrane which effectively changes the inner viscosity. A
Keller-Skalak-type \cite{keller1982} derivation, which consists of a balance
of torque and energy, yields the equations of motion for the angles $\incl$
and $\phase$ \cite{skotheim2007}:
\begin{eqnarray}
  \partial_\tim \incl &=& -\frac{\shearrate}{2}
  -\frac{2\axis_1\axis_2}{\axis_1^2+\axis_2^2}\partial_t \phase +
  \frac{\shearrate}{2}\frac{\axis_1^2-\axis_2^2}{\axis_1^2+\axis_2^2}
  \cos{2\incl} \,,
  \label{eq:skotheim1}
  \\
  \partial_\tim \phase &=& \frac{\shearrate \f_3}{\f_2-\f_1
    \viscin/\viscout}\left(\frac{\elast}{\vol\viscout\shearrate
      \f_3}\sin{2\phase}-\cos{2\incl}\right) \,.
  \label{eq:skotheim2}
\end{eqnarray}
As the underlying equations of motion are overdamped, these equations hold
also for time-dependent shear rate $\shearrate=\shearrate(\tim)$. The
geometrical quantities $\f_i$ depend only upon the semi-axes $\axis_i$ as
given explicitly in Appendix \ref{sec:keller-skalak}.

In the equations of motion (\ref{eq:skotheim1}) and (\ref{eq:skotheim2}) there
are seven independent parameters, namely the hydrodynamic parameters
$\shearrate$, $\viscin$, $\viscout$, an elastic parameter $\elast$ and the
geometric parameters $\axis_1$, $\axis_2$, $\axis_3$ which determine $\vol$,
$\f_1$, $\f_2$, $\f_3$. Three of them can be used to introduce independent
scales.  The volume $V$ defines a length scale, the shear rate $\shearrate$ of
the external flow a time scale, and the elastic energy $\energyscale$ an
energy scale.

\begin{figure}[t]
  \centering
  \begin{minipage}{0.99\linewidth}
    a)
    \begin{center}
      \includegraphics[width=0.60\linewidth]{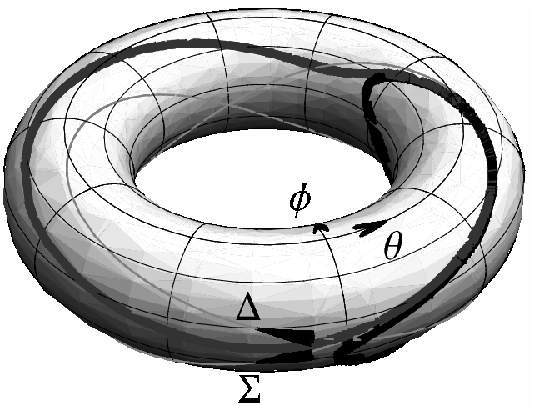}
    \end{center}
  \end{minipage}
  \begin{minipage}{0.99\linewidth}
    b)
    \begin{center}
      \includegraphics[width=0.75\linewidth]{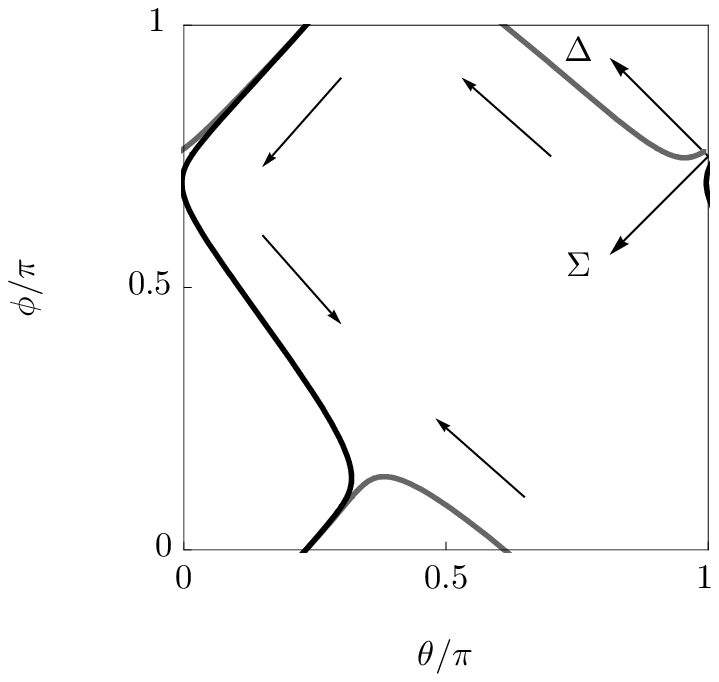}
    \end{center}
  \end{minipage}
  \caption{a) Phase space $T=S^1\times S^1$ of angles $(\incl,\phi)$. b) By
    cutting the torus along the coordinate axes, it can be mapped onto the
    square $[0,\pi]^2$ on the plane by identifying the pair of lines
    $\incl=0$, $\incl=\pi$ and $\phase=0$, $\phase=\pi$. a) and b)
    Coordinates $\incl$, $\phase$, new coordinates $\su$, $\di$ defined by
    eqns.~(\ref{eq:new1}, \ref{eq:new2}), and two typical trajectories (solid
    lines). Arrows in b) denote direction in time.}
  \label{fig:torus}
\end{figure}
Four independent parameters remain, i.e.~the viscosity contrast
$\eta^i/\eta^o$, the ratio between hydrodynamic and elastic energy $(\eta^o
\vol\shearrate)/\elast$, the ratio of the short to the long axis in the shear
plane $\axis_2/\axis_1$, and the ratio of the axis perpendicular to the shear
plane to the long axis $\axis_3/\axis_1$. The first three can be used to
define three equivalent nondimensional parameters, the shifted nondimensional
viscosity contrast
\begin{eqnarray}
  \dimvisc \equiv \frac{\f_1}{-2\f_3} \frac{\viscin}{\viscout} +
  \frac{-\f_2}{-2\f_3}
  \label{eq:dimvisc}
\end{eqnarray}
(note that $\f_1>0$, $\f_2,\f_3<0$, and $\dimvisc>0$ for physical
values), the capillary number
\begin{eqnarray}
  \dimshear \equiv \frac{\vol\viscout(-\f_3)}{\elast} \shearrate \,,
  \label{eq:diminvshear}
\end{eqnarray}
and the eccentricity parameter
\begin{eqnarray}
  \ecc \equiv \arctan{}\frac{\axis_1^2-\axis_2^2}{2\axis_1\axis_2} \,.
  \label{eq:dimecc}
\end{eqnarray}
Here, $\ecc\to0$ corresponds to the spherical case ($\axis_2/\axis_1\to1$),
and $\ecc\to\pi/2$ corresponds to the case $\axis_2/\axis_1\to0$.  It is
convenient to introduce a dimensionless time $\dimtim$ by
\begin{equation}
  d\dimtim \equiv  \frac{2\shearrate}{\dimvisc} d\tim  \,,
\end{equation}
which can also be done in the case of a positive time-dependent shear rate
$\shearrate=\shearrate(t)>0$. We finally arrive at the nondimensional
reformulation of the equations of motion
\begin{eqnarray}
  \partial_\dimtim \incl &=& 
  -\cos\ecc \; \partial_\dimtim \phase - \dimvisc(1-\sin\ecc \cos{2\incl}) \,,
  \label{eq:skotheimscaled1} \\
  \partial_\dimtim \phase &=& 
  -(\diminvshear\sin{2\phase}+\cos{2\incl}) \,,
  \label{eq:skotheimscaled2}
\end{eqnarray}
where $\diminvshear=\diminvshear(\dimtim)>0$ can be time-dependent. The phase
space is the torus $(\incl,\phase)\in T=S^1\times S^1$ (each angle with period
$\pi$, see Fig.~\ref{fig:torus}). Note that the fourth parameter
$\axis_3/\axis_1$ does not enter the nondimensional equations of motion
explicitly.

\subsection{Mean tumbling rate and phase diagram}
\label{sec:phasediagram}
\begin{figure}[t!]
  \centering
  \begin{minipage}{0.04\linewidth}
    a)
  \end{minipage}
  \begin{minipage}{0.94\linewidth}
    \includegraphics[width=0.99\linewidth]{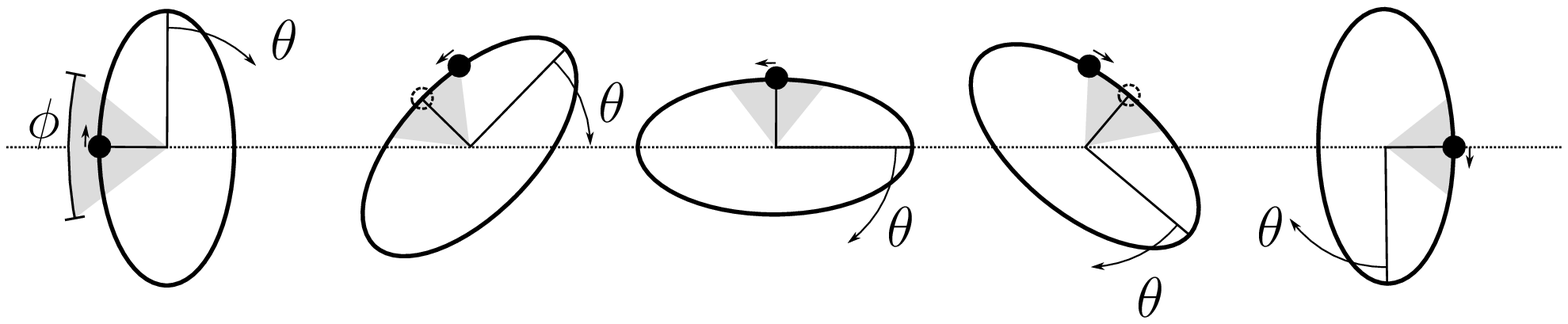}
  \end{minipage}
  \begin{minipage}{0.04\linewidth}
    b)
  \end{minipage}
  \begin{minipage}{0.94\linewidth}
    \includegraphics[width=0.99\linewidth]{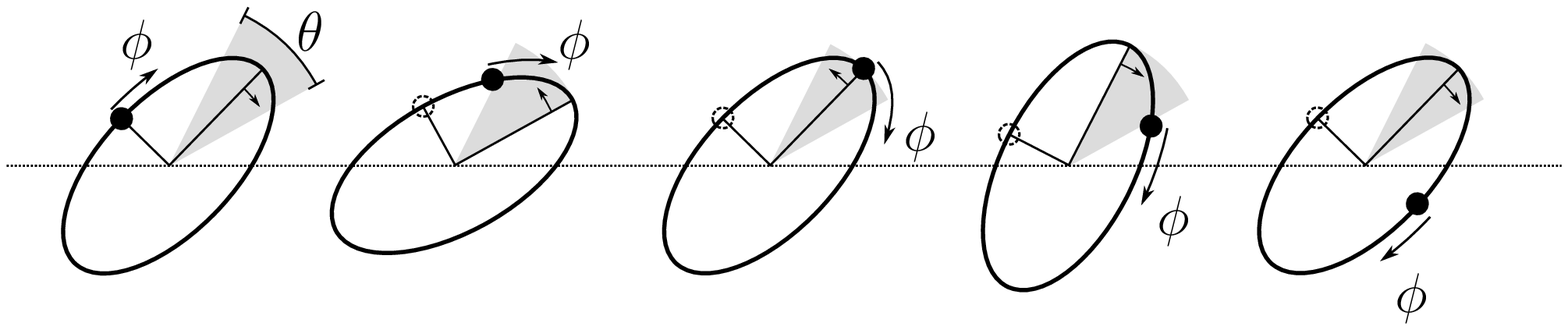}
  \end{minipage}
  \begin{minipage}{0.04\linewidth}
    c)
  \end{minipage}
  \begin{minipage}{0.94\linewidth}
    \includegraphics[width=0.99\linewidth]{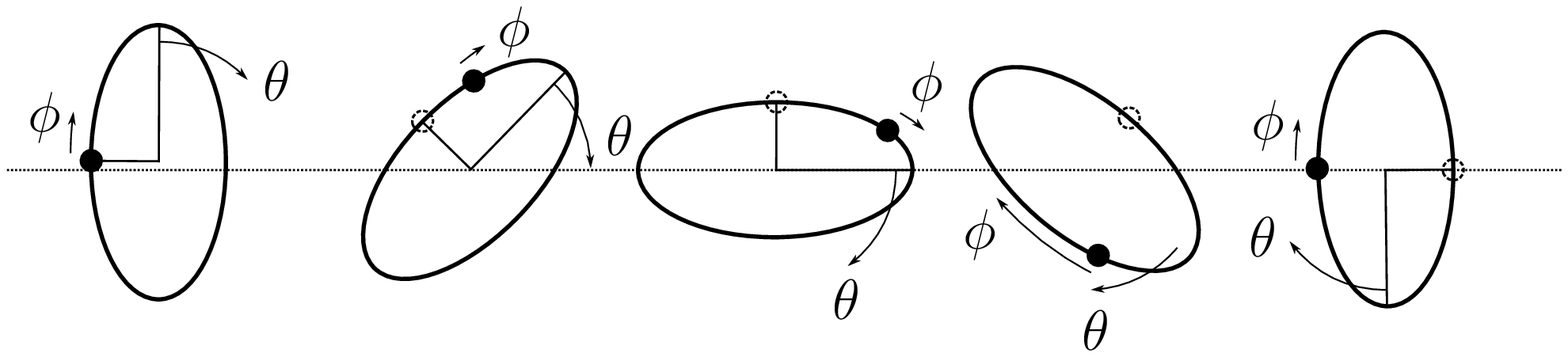}
  \end{minipage}
  \begin{minipage}{0.99\linewidth}
    d)
    \begin{center}
      \includegraphics[width=0.7\linewidth]{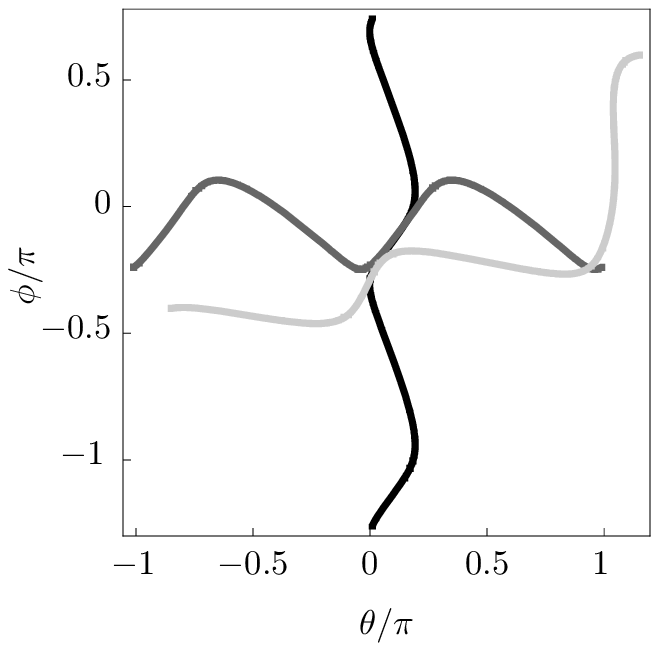}
    \end{center}
  \end{minipage}
  \caption{a) Tumbling: The capsule flips with montonously changing
    inclination angle $\incl$ and an oscillating phase angle $\phase$. b)
    Swinging: The inclination angle $\incl$ oscillates around a constant value
    while the phase angle is changing monotonously. c) Intermittent motion:
    Both angles grow without bounds leading to a mixture of tumbling and
    swinging motion. d) Trajectories for swinging (black), tumbling (dark
    grey), and intermittent motion (light grey).
  }
  \label{fig:tttu}
\end{figure}
\begin{figure*}[t!]
  \centering
  \begin{minipage}{0.3\linewidth}
    a)\\
    \includegraphics[width=0.99\linewidth]{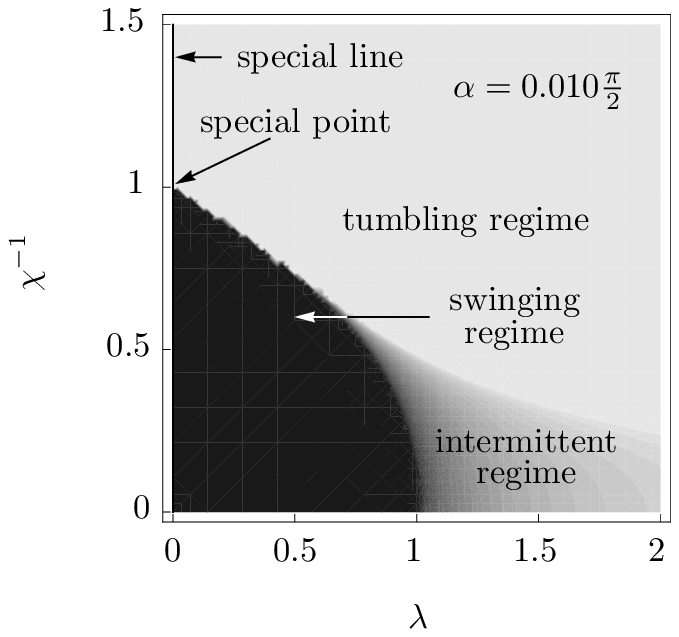}
  \end{minipage}
  \begin{minipage}{0.3\linewidth}
    b)\\
    \includegraphics[width=0.99\linewidth]{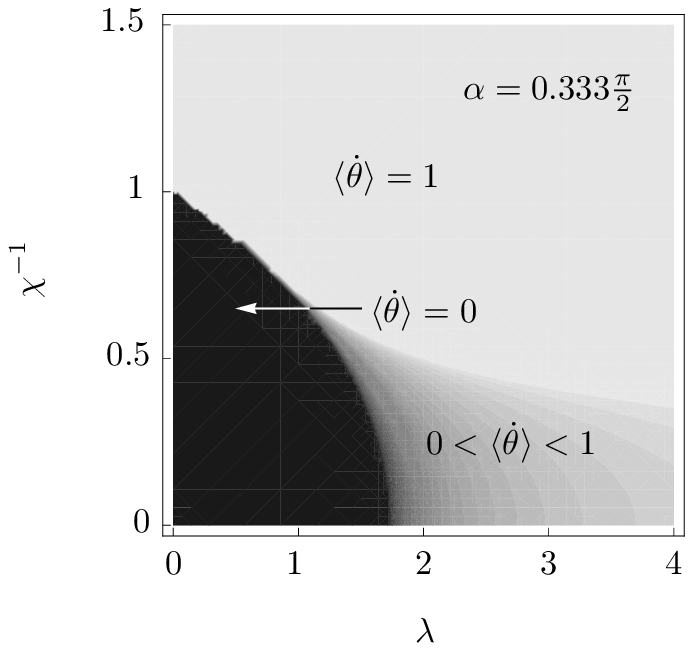}
  \end{minipage}
  \begin{minipage}{0.3\linewidth}
    c)\\
    \includegraphics[width=0.99\linewidth]{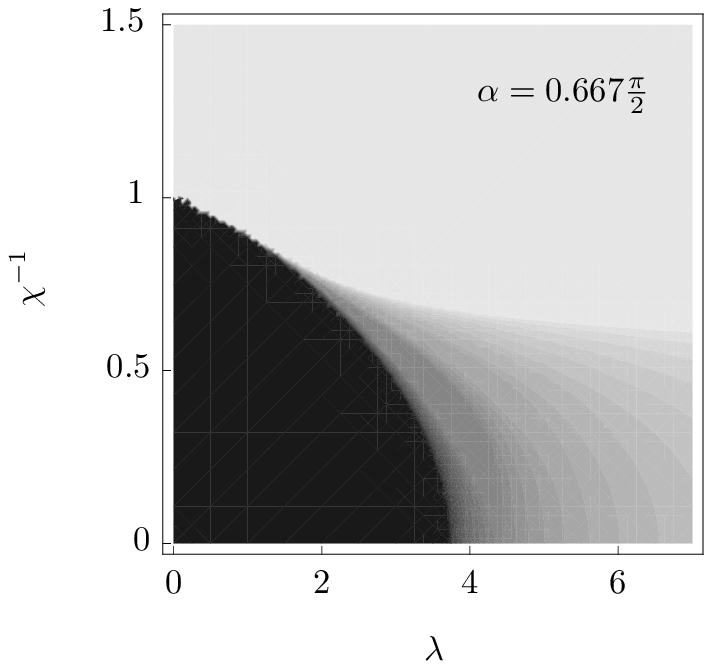}
  \end{minipage}
  \caption{Typical phase diagrams for different eccentricity parameters
    $\ecc=0.010\frac{\pi}{2}$, $0.333\frac{\pi}{2}$, $0.667\frac{\pi}{2}$ with
    the dimensionless inverse capillary number $\diminvshear$ and the
    dimensionless viscosity ratio $\dimvisc$ as axes. The mode of motion for
    each point in the phase diagram and thereby the transition lines between
    swinging (black), intermittent (grey shading) and tumbling regime (light
    grey) can be distinguished by the mean tumbling rate $\tumb$ defined by
    eqns.~(\ref{eq:tumb}-\ref{eq:mean2}) which is proportional to the color
    shading. In panel a) the special line $\dimvisc=0$ and the special point
    at $\dimvisc=0$ and $\diminvshear=1$ which are discussed in section
    \ref{sec:constant-shear-rate} are depicted explicitly.}
  \label{fig:dimphase}
\end{figure*}

The solutions of the equations (\ref{eq:skotheimscaled1}) and
(\ref{eq:skotheimscaled2}) in the case of a time-constant shear flow have been
examined in Ref.~\cite{skotheim2007}. The phase diagram consists of three
different regimes depending upon the value of three parameters $\dimvisc$,
$\diminvshear$, and $\ecc$.  As shown in Fig.~\ref{fig:tttu}, the capsule can
either tumble with a monotonously changing inclination angle $\incl$ and an
oscillating phase angle $\phase$ or tank-tread with an oscillating inclination
angle $\incl$ and a monotonously changing phase angle $\phase$, a motion
called swinging.  There is a third regime in between, where the capsule both
tumbles and tank-treads (either successivly or simultanously), which is called
``intermittent'' regime in Ref.~\cite{skotheim2007} (see
Fig.~\ref{fig:dimphase}). For low shear rates, i.e.~for large values of
$\diminvshear$, the hydrodynamic flow is too weak to overcome the elastic
barrier $\elast$ due to the shape memory. Thus, the capsule tumbles for large
$\diminvshear$. For smaller values of $\diminvshear$, i.e.~for higher shear
rates, the transition to the intermittent or swinging regime occurs.

The dynamics in the reduced model can conveniently be characterized by
investigating the normalized mean tumbling rate
\begin{eqnarray}
  \tumb
  &\equiv&
  \frac{\mean{\partial_\dimtim \incl}}
  {\mean{\partial_\dimtim \incl}+\mean{\partial_\dimtim \phase}} 
  \label{eq:tumb}
\end{eqnarray}
as an order parameter. Here the mean rates of inclination and phase angle are
given by
\begin{eqnarray}
  \mean{\partial_\dimtim \incl} &\equiv &
  \lim\limits_{T\to\infty}\frac{1}{T}
  \int\limits_0^T d\dimtim \partial_\dimtim \incl(\dimtim)
  ~~\text{and} 
  \label{eq:mean1}
  \\
  \mean{\partial_\dimtim \phase} &\equiv &
  \lim\limits_{T\to\infty}\frac{1}{T}
  \int\limits_0^T d\dimtim \partial_\dimtim \phase(\dimtim) \,,
  \label{eq:mean2}
\end{eqnarray}
respectively.  In a stable tumbling motion, the inclination angle grows
without bounds while the phase angle oscillates, which implies
$\mean{\partial_\dimtim \phase}/\mean{\partial_\dimtim \incl} = 0$ and a mean
tumbling rate $\tumb = 1$ in the long time limit. Conversely, in a stable
swinging motion, the phase angle grows without bounds while the inclination
angle oscillates, which implies $\mean{\partial_\dimtim
  \incl}/\mean{\partial_\dimtim \phase} = 0$ in the long time limit thus
$\tumb = 0$. In the intermittent regime, the mean tumbling rate takes values
between $0$ and $1$. Typical phase diagrams showing grey scale plots of the
mean tumbling rate $\tumb$ as obtained from solving equations
(\ref{eq:skotheimscaled1}) and (\ref{eq:skotheimscaled2}) numerically can be
seen in Fig.~\ref{fig:dimphase}. Here, the axes consist of the dimensionless
viscosity ratio $\dimvisc$ and the inverse capillary number $\diminvshear$,
while the eccentricity $\ecc$ is constant.

Even though it is not central for this paper, we note for completeness that
the status of the intermittent regime is still under debatte. Kessler et
al.~\cite{kessler2007} and Sui et al.~\cite{sui2008a} solved the full dynamics
of a 3d elastic capsule using a spectral method and a immersed boundary
lattice Boltzmann method, respectively.  While the reduced model captures the
swinging and tumbling regime semi-quantitatively compared to fully numerical
studies \cite{kessler2007}, the intermittent regime has been contested as an
artifact of the reduced model. Neither study found any indications of
intermittency, but rather a transition towards swinging. There was also no
direct evidence of intermittent motion in experiments \cite{abkarian2007}.

\subsection{Quasi-spherical case}
\label{sec:quasi-spherical-case}

Since the phase diagram is qualitatively similiar for all small values of
$\ecc \lesssim 1$ (see Fig.~\ref{fig:dimphase}), it is sufficient to
investigate the quasi-spherical case, for which analytical progress becomes
possible. We set $\axis_2=(1-\sm)\axis_1$ where $\sm\ll1$ is a small parameter
and assume the difference of $\axis_3-\axis_1$ to be also of order $\sm$. The
three dimensionless parameters introduced above then depend on $\sm$ to first
order as (see Appendix \ref{sec:notation} for definition of symbols)
\begin{eqnarray}
  \ecc&\approx&\sm \,,
  \label{eq:exp1}
  \\
  \dimvisc &\approx& 
  \frac{3+2\viscin/\viscout}{5} \sm \,,
  \label{eq:exp2}
  \\
  \dimshear &\approx& 
  \frac{5\vol\viscout}{\elast}\shearrate \sm \,.
\end{eqnarray}
In the quasi-spherical case, the elastic energy can be calculated for any
elastic model. In the regime of small deformations the elastic energy scales
quadratic with the eccentricity
\begin{eqnarray}
  \elast \equiv \elastprop \sm^2 
\end{eqnarray}
with $\elastprop\sim1$, leading to
\begin{eqnarray}
  \dimshear &\approx& 
  \frac{5\vol\viscout}{\elastprop}\shearrate 
  \sm^{-1} \,.
  \label{eq:exp3}
\end{eqnarray}
For given values of all physical parameters, leaving aside the shear rate
$\shearrate$ and the small parameter $\sm$, the pre-factors in the above
expansions (\ref{eq:exp1}), (\ref{eq:exp2}), and (\ref{eq:exp3}) are of the
order of unity in the quasi-spherical limit $\sm\to0$.  In this case, we are
restricted to a small left hand stripe in the phase diagram (see
Fig.~\ref{fig:dimphase}) defined by $\dimvisc\sim\sm$. Here, the disputed
intermittent regime has no influence on the dynamics and can be ignored. Since
the transition between tumbling and swinging takes place at
$\diminvshear\sim1$, we will later specify the shear rate to be of the order
of the expansion parameter $\shearrate\sim\sm$. For the following expansion we
merely require the scaling of $\diminvshear$ not to be smaller than $\sm^1$.

Before we expand the equations of motion (\ref{eq:skotheimscaled1}) and
(\ref{eq:skotheimscaled2}) in $\sm$, we introduce a suitable stretched,
rotated, and translated frame in the $\incl\phase$-plane (see
Fig.~\ref{fig:torus}) with coordinates
\begin{eqnarray}
  \su &\equiv& -\left(\phase+\incl+\frac\pi4\right) \,, 
  \label{eq:new1} \\
  \di &\equiv& \phase-\incl+\frac\pi4 \,,
  \label{eq:new2}
\end{eqnarray}
where $\su$ is, up to a constant, the angle of a tracer particle with respect
to the direction of the shear flow. In these coordinates, the mean tumbling
rate (\ref{eq:tumb}) can be written as
\begin{equation}
  \tumb = \frac{
    \mean{\partial_\dimtim \di}+\mean{\partial_\dimtim \su}
  }{
    2\mean{\partial_\dimtim \su}
  } 
  = \frac 1 2 \left(1 +
    \frac{
      \mean{\partial_\dimtim \di}
    }{
      \mean{\partial_\dimtim \su}} \right)
  \,.
  \label{eq:tumb2}
\end{equation}
Finally, expansion of the equations of motion (\ref{eq:skotheimscaled1}) and
(\ref{eq:skotheimscaled2}) up to first order in $\sm$ (note that
$\dimvisc\sim\sm$, $\ecc\approx\sm$) yields the quasi-spherical equations of
motion in the new coordinates
\begin{eqnarray}
  \partial_\dimtim \su &=& \dimvisc \,, 
  \label{eq:quasi1} \\
  \partial_\dimtim \di &=& 4 \sin\su\sin\di 
  + \dimvisc + 2(\diminvshear-1) \cos{(\su-\di)} \,.~~~
  \label{eq:quasi2}
\end{eqnarray}
With the initial condition $\su_0\equiv\su(0)$ the solution of the first
equation is
\begin{equation}
  \su(\dimtim) = \su_0 + \dimvisc\dimtim \,,
\end{equation}
i.e.~a tracer particle moves with constant angular velocity with respect to
the dimensionless time $\dimtim$. Assuming that the shear rate does not change
sign, we can use the angle $\su$ as a time quantity to arrive at an autonomous
differential equation on the torus
\begin{eqnarray}
  \dimvisc \frac{d \di}{d \su} 
   = 4 \sin\su\sin\di 
       + \dimvisc + 2(\diminvshear-1) \cos{(\su-\di)} \,,~~
      \label{eq:autonomous}
\end{eqnarray}
where $\diminvshear$ is now considered to be a function of $\su$ rather than
$\dimtim$. Here, $\su$ and $\di$ are not restriced to the interval $[0,\pi)$
but can take all real values, thereby accounting for the number of revolutions
on the torus.

\section{Constant shear rate}
\label{sec:constant-shear-rate}

We first summarize the big picture derived below in detail for the shape
dynamics expressed by $\su$ and $\di$ close to the special line $\dimvisc=0$
(see Fig.~\ref{fig:dimphase}), corresponding to the quasi-spherical limit
$\sm\to0$ at fixed inverse capillary number $\diminvshear$. For a systematic
expansion in $\dimvisc$, we first investigate the dynamics at the special
line. We find a set of fixed points connected by curves, of which some are
stable, some unstable, and some neutral (see Fig.~\ref{fig:fixedpoints}). For
capillary numbers $\diminvshear\neq1$, closed and separated curves in the
phase space exist which consist solely of either stable or unstable fixed
points. Upon perturbation to first order in $\dimvisc$ these lines turn to
stable or unstable limit cycles, as the perturbation does not alter the
absolute stability. These limit cycles correspond to swinging and tumbling for
$\diminvshear<1$ and $\diminvshear>1$, respectively. At the special point
$\diminvshear=1$, the lines of fixed points cross. Here, all vertical lines of
fixed points consist solely of neutral fixed points, whereas all horizontal
lines of fixed points consist of segments of either stable or unstable fixed
points.  Upon perturbation, the neutral fixed points can become either stable
or unstable. Therefore the dynamics of the system close to the special point
needs to be studied more carefully to first order in $\dimvisc$. Analytic
determination of the resulting limit cycles becomes possible by considering
the trajectories close to the stable, unstable, and neutral fixed points
separately and joining these with the method of matched asymptotic expansion.

\subsection{Zero order expansion on special line}
\label{sec:special-line}

\begin{figure*}[t]
  \centering
  \begin{minipage}{0.3\linewidth}
  a)\\\includegraphics[width=0.99\linewidth]{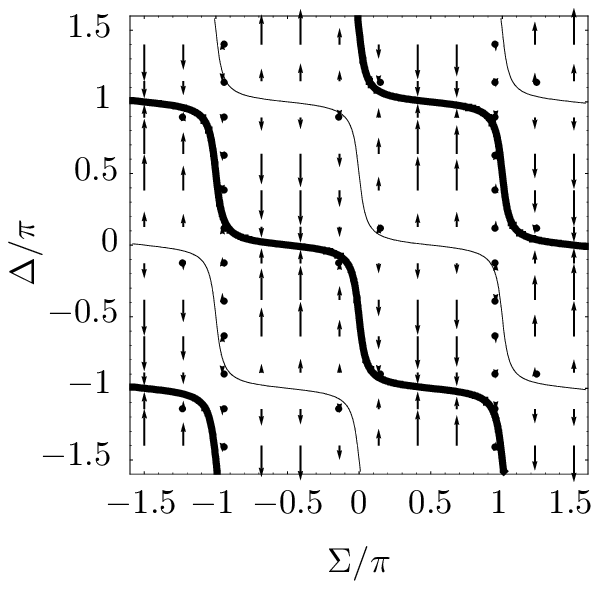}
  \end{minipage}
  \begin{minipage}{0.3\linewidth}
  b)\\\includegraphics[width=0.99\linewidth]{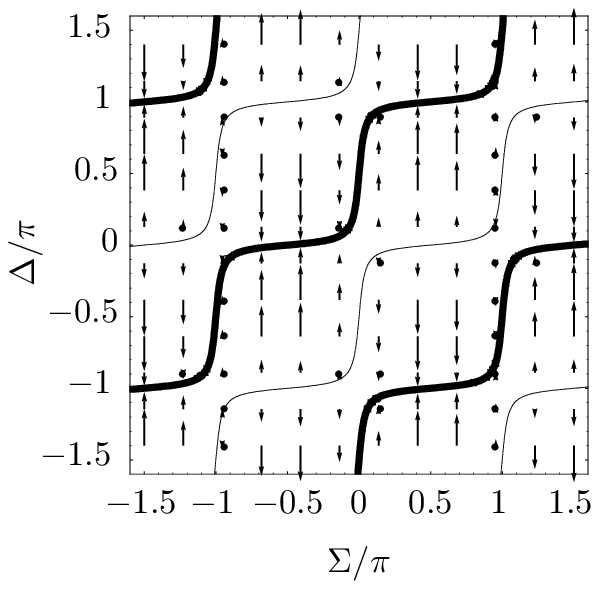}
  \end{minipage}
  \begin{minipage}{0.3\linewidth}
  c)\\\includegraphics[width=0.99\linewidth]{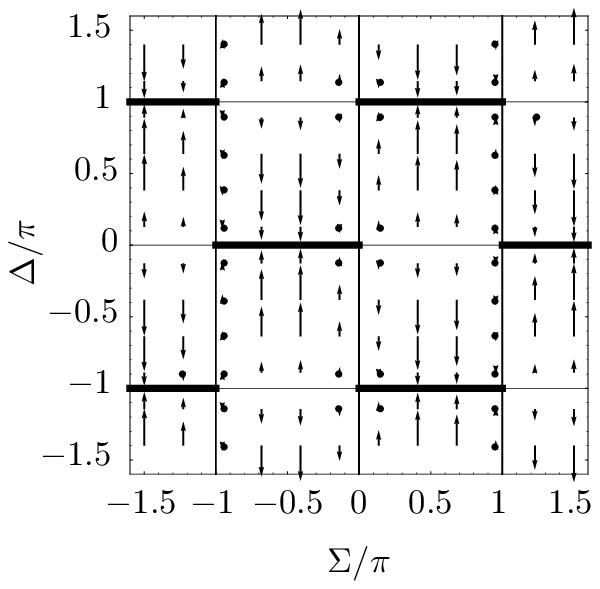}
  \end{minipage}
  \caption{Vector field and curves of connected fixed points on the special
    line $\dimvisc=0$ for $\ecc\to1$, and a) $\diminvshear=0.8<1$, b)
    $\diminvshear=1.2>1$, and c) $\diminvshear=1$ (checkerboard
    pattern). Thick lines correspond to stable fixed points, thin lines to
    unstable fixed points, and regular lines to neutral fixed points.}
  \label{fig:fixedpoints}
\end{figure*}

We start by investigating the special line $\dimvisc=0$, where the equations
of motion
\begin{eqnarray}
  \partial_\dimtim \su &=& 0 \,, 
  \label{eq:eqn1}
  \\
  \partial_\dimtim \di &=& 4 \sin\su\sin\di 
  + 2(\diminvshear-1) \cos{(\su-\di)}
  \label{eq:eqn2}
\end{eqnarray}
immediately lead to time-constant $\su$. For all values of $\diminvshear$
there are two connected lines of fixed points (see
Fig.~\ref{fig:fixedpoints}) which bifurcate into limit cycles upon a
perturbation with $\dimvisc>0$. The position and stability character of these
lines of fixed points depend on the inverse capillary number
$\diminvshear$. The corresponding regimes on the special line $\dimvisc=0$ are
seperated by the special point ($\diminvshear=1$, $\dimvisc=0$).

At this special point, there are two straight lines of connected fixed points,
namely $\su=k\pi$ with arbitrary $\di$ and $\di=k\pi$ with arbitrary $\su$ and
integer $k$. These lines of fixed points build up a checkerboard pattern as
can be seen in Figs.~\ref{fig:fixedpoints} c) and \ref{fig:fixedpoints2}.
We now discuss the stability of each fixed point. The eigenvalue of a
linearization around a fixed point in the direction of the connected fixed
points is always zero. The sign of the other eigenvalue determines the
stability of the fixed point in the perpendicular direction. Fixed points with
a positive, negative or zero eigenvalue are called unstable, stable or
neutral, respectively. The vertical lines of fixed points are neutral as $\su$
is constant (\ref{eq:eqn1}), while stable and unstable segments alternate on
the horizontal lines as can be seen in Fig.~\ref{fig:fixedpoints}~c), where
the vector field corresponding to the equations of motion is shown. Thus, each
square consists of two opposite neutral sides and two opposite lines of which
one is stable and one is unstable.

For arbitrary points on the special line $\dimvisc=0$ with
$\diminvshear\neq1$, there are two separated curves of fixed points (see
Fig.~\ref{fig:fixedpoints} a, b) separated from the checkerboard pattern by a
distance of order $\abs{\diminvshear-1}^{1/2}$ as can be calculated using
eqn.~(\ref{eq:eqn2}). Here, the fixed points on one single curve are
either all stable or unstable.

\subsection{First order expansion away from the transition}
\label{sec:first-order}

Since the system moves with constant velocity $\dimvisc$ along the
$\su$-direction (see eqn.~(\ref{eq:quasi1})), fixed points exist only for
vanishing viscosity contrast $\dimvisc=0$. For finite $\dimvisc>0$, the lines
of fixed points turn into limit cycles. Excluding the region close to the
special point (i.e.~excluding $\diminvshear-1\sim \dimvisc$), the perturbation
due to finite $\dimvisc>0$ is too small to change the stability
qualitatively. Thus, the stable character of the original stable line of fixed
points as well as its topology remain unchanged. The stable limit cycle for
small $\lambda>0$ and $\diminvshear<1$ leads to a decreasing $\di$, while
$\su$ is increasing (see Fig.~\ref{fig:fixedpoints}~a). In the long time limit
the mean rates have the same magnitude $\mean{\partial_\dimtim
  \di}=-\mean{\partial_\dimtim \su}$, resulting in a swinging motion with
vanishing mean tumbling rate $\tumb=0$ (see eqn.~(\ref{eq:tumb2})). For
$\diminvshear>1$ the stable limit cycle leads to $\mean{\partial_\dimtim
  \di}=\mean{\partial_\dimtim \su}>0$ (see Fig.~\ref{fig:fixedpoints}~b),
resulting in a tumbling motion with mean tumbling rate $\tumb=1$ (see
eqn.~(\ref{eq:tumb2})). This is consistent with the phase diagrams shown in
Fig.~\ref{fig:dimphase}.

\subsection{Matched asymptotic expansion close to the transition}
\label{sec:matched-asymptotics}

We now investigate the system for a finite but small viscosity contrast
$\dimvisc\sim\sm$ close to the special point ($\diminvshear=1$, $\dimvisc=0$),
where we specialize to straight lines emerging from the special point. We
define a slope parameter $\slope\sim1$ of order unity by
\begin{eqnarray}
  \diminvshear \equiv 1 + \frac{\slope-1}{2}\dimvisc
  \label{eq:slope-def}
\end{eqnarray}
and solve the autonomous equation of motion (\ref{eq:autonomous}) 
\begin{eqnarray}
  \frac{d \di}{d \su} 
   &=& \frac 4 \dimvisc \sin\su\sin\di \\
      && + \slope + (\slope-1)(\cos{(\su-\di)}-1) \nonumber
  \label{eq:central-asymptotic}
\end{eqnarray}
asymptotically.
\begin{figure}[t]
  \centering
  \includegraphics[width=0.80\linewidth]{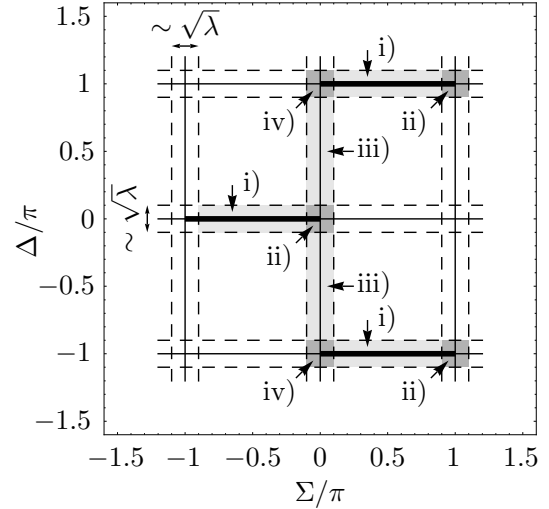}
  \caption{The four regions i) to iv) with linear dimension of order
    $\sqrt\dimvisc$ in which asymptotic solutions are obtained in
    section~\ref{sec:matched-asymptotics} for $0<\dimvisc\ll1$. i) Light grey
    region: thick stable line. ii) Dark grey region: junction region, crossing
    of stable and neutral line. iii) Light grey region: neutral line. iv) Dark
    grey region: crossing of neutral and stable line.}
  \label{fig:fixedpoints2}
\end{figure}
\begin{figure*}[t]
  \begin{minipage}{0.3\linewidth}
  i)\\\includegraphics[width=0.99\linewidth]{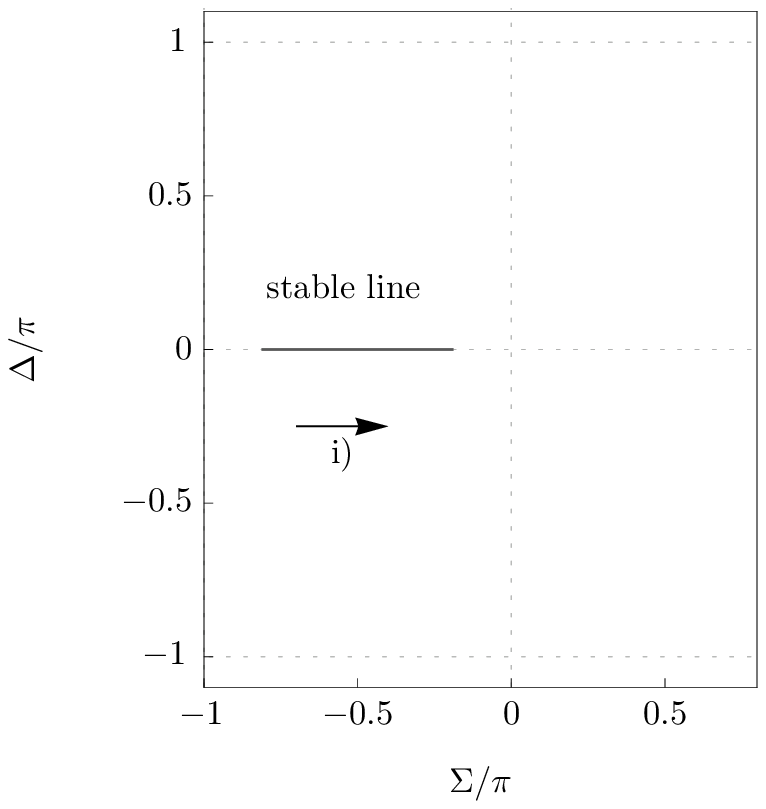}  
  \\iv)\\\includegraphics[width=0.99\linewidth]{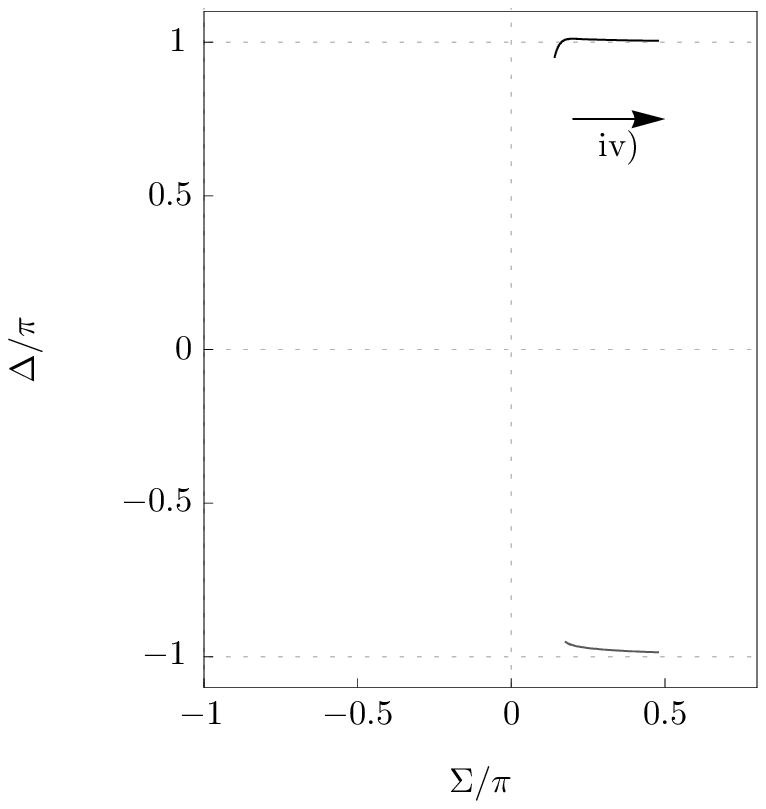}  
  \end{minipage}
  \begin{minipage}{0.3\linewidth}
  ii)\\\includegraphics[width=0.99\linewidth]{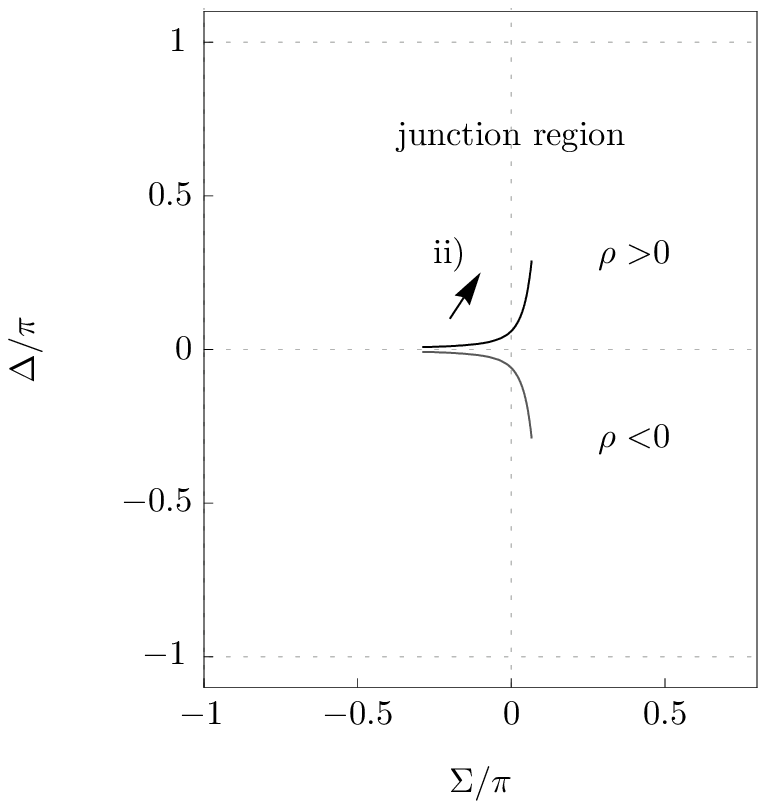}  
  \\v)\\\includegraphics[width=0.99\linewidth]{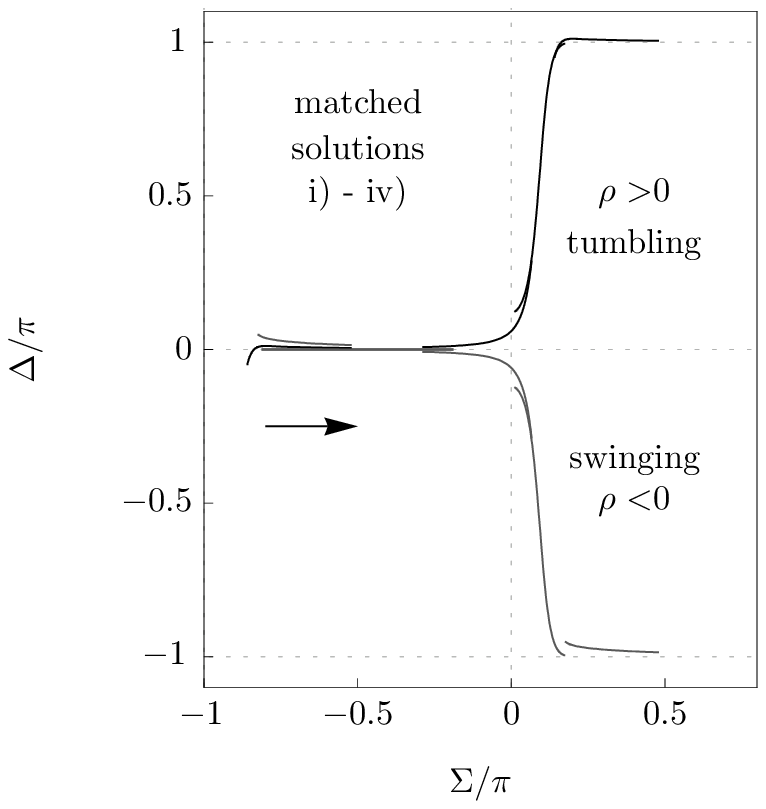}  
  \end{minipage}
  \begin{minipage}{0.3\linewidth}
  iii)\\\includegraphics[width=0.99\linewidth]{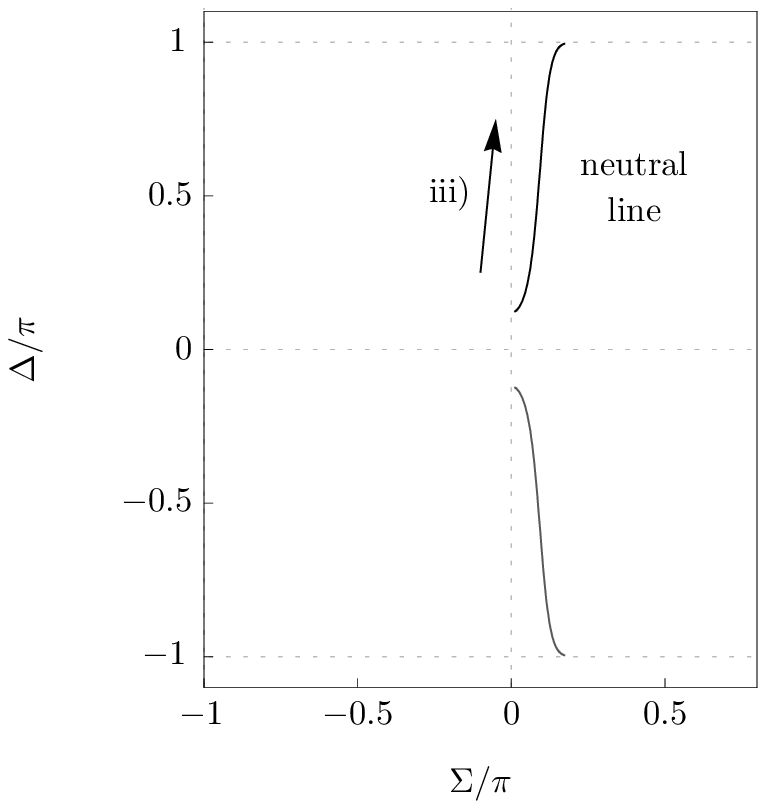}  
  \\vi)\\\includegraphics[width=0.99\linewidth]{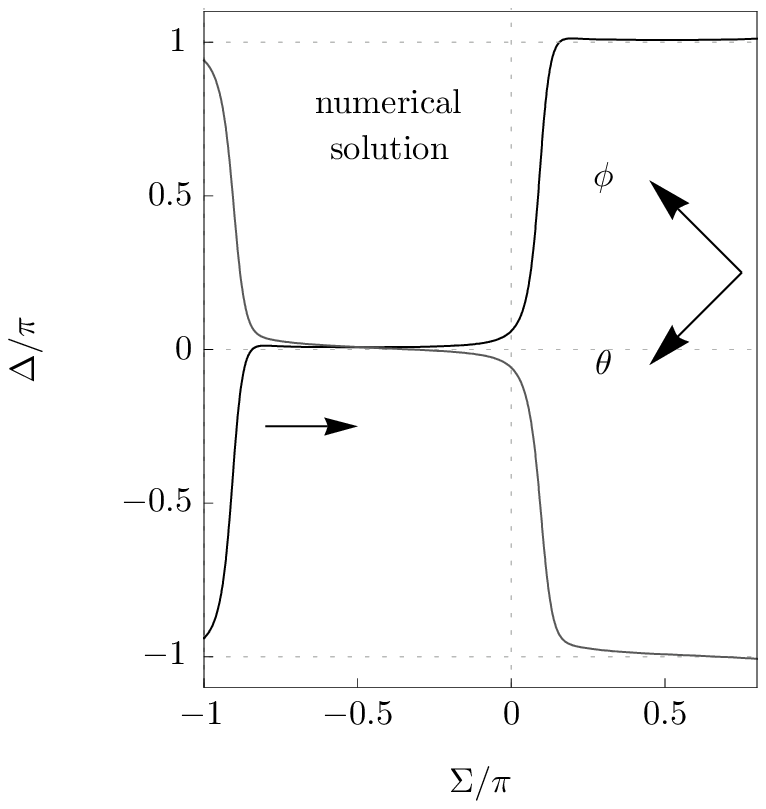}  
  \end{minipage}
  \caption{i) to iv) Asymptotically matched solutions in the four regions
    close to the stable and neutral lines. v) All asymptotic solutions
    combined. vi) Numerical solution obtained by direct integration of the
    full equations of motion. Parameters are chosen from the tumbling
    regime (black curves) with $\ecc=0.001\pi$, $\dimvisc=0.3$, $\slope=1$ and
    from the swinging regime (grey curves) with $\slope=-1$. The arrows denote
    the direction in time.}
  \label{fig:asymptotic}
\end{figure*}
Since the curves of fixed points are separated by the lines of fixed points of
the checkerboard pattern by approximately $\sqrt{\abs{\diminvshear-1}} \sim
\sqrt\dimvisc$, the stable limit cycle should be within stripes of width
$\sqrt\dimvisc$ from the lines of fixed points of the special point. In each
of the four regions i) to iv) shown in Fig.~\ref{fig:fixedpoints2}, we
expand the equations of motion up to lowest order in $\dimvisc\sim\sm$ and
solve them analytically. By the method of asymptotic matching \cite{hinch1991}
the constants of integration can be deduced step by step. Here, we proceed as
follows: We start with the general solution of region i) and match the
solution of region ii). We continue by matching region ii) with region iii)
and region iii) with region iv). Finally, we match the solution of region iv)
with the general solution of region i) to obtain a closed trajectory on the
torus. Thus, we arrive at the unique analytical solution of the stable limit
cycle.

We now give a brief description of the solution which should be read while
comparing Fig.~\ref{fig:asymptotic}, which shows typical graphs of the
obtained solutions. For a more detailed derivation, which includes the
expanded equations of motions and the asymptotically matched solutions, we
refer the reader to Appendix \ref{sec:matching}.
\begin{itemize}
\item[i)] We start with the region close to the stable line with $\su\sim1$
  and $\di\sim\sqrt\dimvisc$ (see Fig.~\ref{fig:fixedpoints2}). As is shown in
  Appendix \ref{sec:matching}, in region i) the limit cycle is to first order
  in $\dimvisc$ simply given by the original stable line (see
  eqn.~(\ref{eq:region1}) and Fig.~\ref{fig:asymptotic} i). Thus, the system
  runs on the stable line $\di(\su)=0$ with $-\pi<\su<0$ irrespective of the
  slope parameter $\slope$. When the system starts in the vicinity of the
  limit cycle, it relaxes quickly to the stable line.
\item[ii)] In the region $\su\sim\sqrt\dimvisc$ and $\di\sim\sqrt\dimvisc$,
  where stable and neutral line meet, the vector field corresponding to
  $\dimvisc=0$ is small enough for the finite but small value of $\dimvisc$ to
  have a significant influence on the vector field and thus on the
  motion. Here, the exact value of the shear rate or slope parameter $\slope$
  is critical as can be seen by the matched solution (\ref{eq:region2}) whose
  sign in the long-time limit depends only on the sign of the slope~$\slope$
  \begin{eqnarray}
    \lim\limits_{\su\to\infty}\sign{(\di(\su))} &=& \sign{\slope} \,.
  \end{eqnarray}
  A typical graph can be seen in Fig.~\ref{fig:asymptotic} ii) for both
  cases $\slope<0$ and $\slope>0$. For a negative slope $\slope<0$ the neutral
  line with $\di<0$ is choosen which leads to a swinging motion. Conversely,
  for a positive slope $\slope>0$ the neutral line with $\di>0$ is choosen,
  corresponding to a tumbling motion (see region iii)). Region ii) with
  $\su\sim\di\sim\sqrt{\dimvisc}$ acts as a junction which only depends on the
  sign of the slope parameter $\slope$.
\item[iii)] In the region $\su\sim\sqrt\dimvisc$ and $\di\sim1$ close to the
  neutral line, the matched solution (\ref{eq:region3}) describes the
  relaxation towards the next stable line (see Fig.~\ref{fig:asymptotic} iii),
  which has been chosen in region ii).
\item[iv)] In the region $\su\sim\sqrt\dimvisc$ and
  $\di\pm\pi\sim\sqrt\dimvisc$, where neutral and unstable lines meet again,
  the matched solution (\ref{eq:region4}) describes a relaxation toward the
  stable line for all values of $\slope$ (see Fig.~\ref{fig:asymptotic}
  iv). The system then starts over again in region i) close to the stable
  line.
\end{itemize}
Panels v) and vi) of Fig.~\ref{fig:asymptotic} show a comparison of the
matched asymptotic solutions and the numerically computated stable limit
cycle, with excellent agreement.

Summarizing the dynamics, the system starts running along a horizontal stable
line $\di=0$. At its end $\su\simeq0$, it chooses one side depending on the
sign of $\slope$ and runs close to the vertical neutral line towards the
neighbouring horizontal stable line. For negative $\slope<0$, the angle $\di$
is decreasing along the neutral line. This case corresponds to a motion with
oscillating inclination angle $\incl$ and monotonously decreasing phase angle
$\phase$ (see Fig.~\ref{fig:asymptotic} vi), resulting in a swinging
motion. For positive $\slope>0$, the angle $\di$ is increasing along the
neutral line. This case corresponds to a motion with oscillating phase angle
$\phase$ and monotonously decreasing inclination angle $\incl$, resulting in a
tumbling motion.

In summary, these results imply for the phase diagram that the boundary
between the tumbling and swinging regime is given by the line
$\diminvshear=1-\dimvisc/2$, which corresponds to the critical value
$\slope=0$, in first order in $\sm$.

\section{Time-modulated  shear rate}
\label{sec:time-modul-shear}

\begin{figure*}[t]
  \centering
  \begin{minipage}{0.49\linewidth}
    a)\\\includegraphics[width=0.99\linewidth]{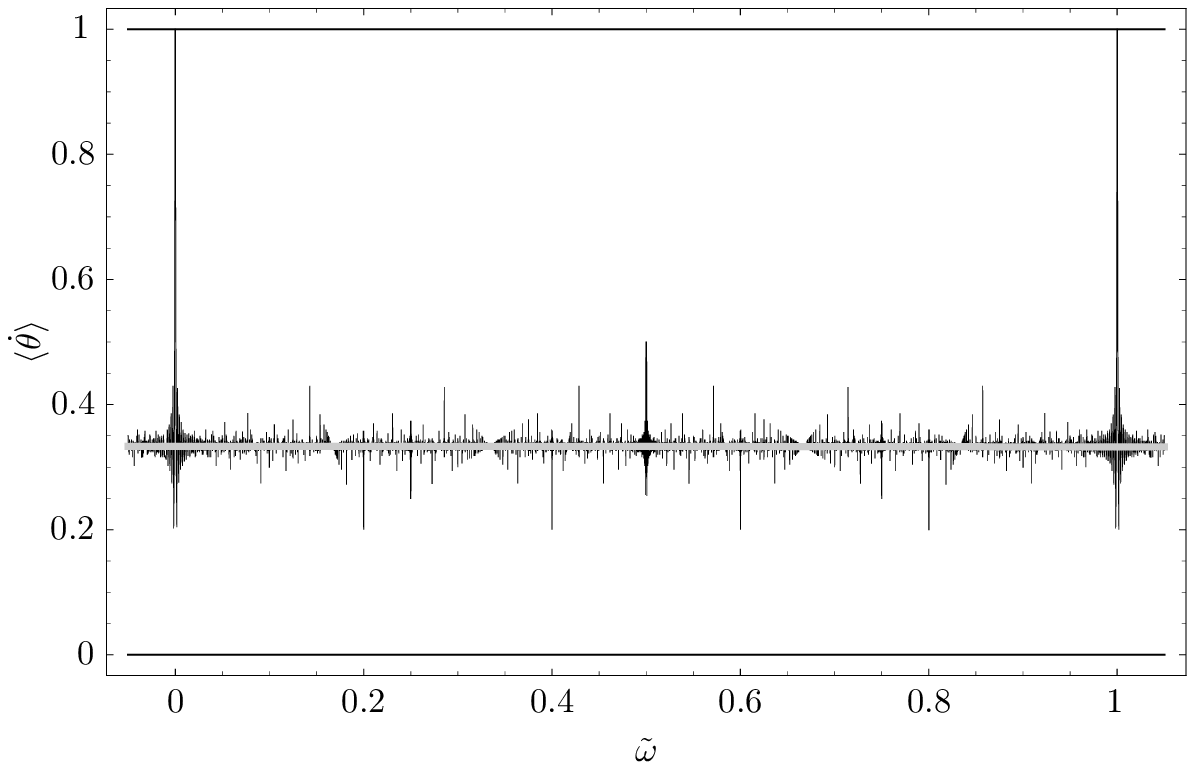}
  \end{minipage}
  \begin{minipage}{0.49\linewidth}
    b)\\\includegraphics[width=0.99\linewidth]{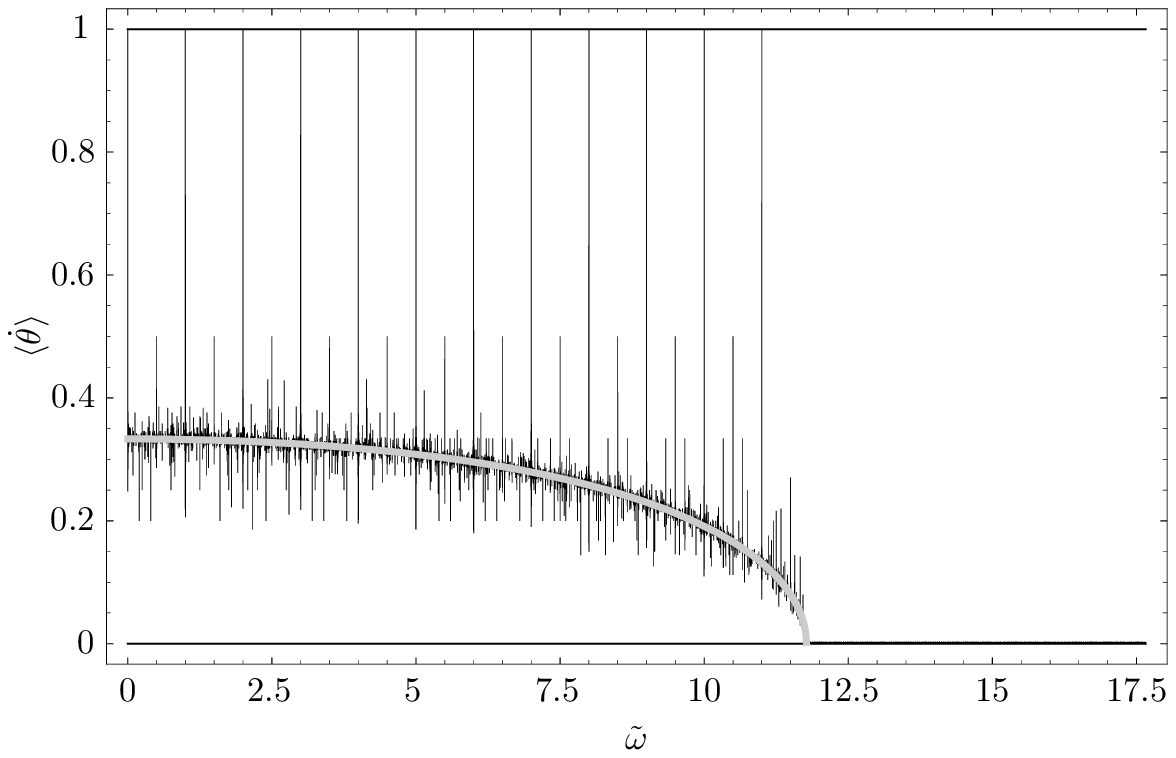}
  \end{minipage}
  \caption{Typical numerical value of the mean tumbling rate $\tumb$ as a
    function of $\dimfreq$ in finite simulation runs of time-dependent shear
    flow. a) Low frequencies $0\leq\dimfreq\leq1$. b) Whole spectrum. -- There
    are several resonance peaks of different height whose width depends upon
    both the length of the simulation run and the amplitude of the oscillating
    shear rate. The smooth background starts at a constant value for small
    frequencies and finally vanishes at a cut-off frequency. The thick grey
    line is an analytical result (\ref{eq:background}) describing the
    background with excellent agreement. Parameters (see
    eqns.~(\ref{eq:osci1}), (\ref{eq:osci2}) and
    sec.~\ref{sec:harm-oscill-shear}): $\ecc=0.001\pi$, $\dimvisc=0.01$,
    $\slope_0=-1$, $\slope_1=2$, $\ph_0=0$, $\su_f=500\pi$.}
  \label{fig:resonancenum}
\end{figure*}
We now investigate the dynamics of a capsule in a time-modulated shear flow
and specialize to a periodically oscillating positive shear rate
$\shearrate(\tim)>0$ with period $\period$, frequency
$\freq\equiv2\pi/\period$ and mean value
\begin{eqnarray}
  \shearrate_0\equiv\frac 1 \period \int\limits_0^\period\shearrate(t)dt \,. 
\end{eqnarray}
Then the slope parameter $\slope(\su)$ defined by
equations~(\ref{eq:diminvshear}), (\ref{eq:slope-def}), and (\ref{eq:quasi1})
is periodic and can be written as
\begin{equation}
  \slope(\su) = \slope_0 + \slope_1\amp(\su) \,.
  \label{eq:osci1}
\end{equation}
Here, the oscillatory function $\amp(\su)$ has a vanishing mean value
\begin{eqnarray}
  \int\limits_0^{\pi/\dimfreq}\amp(\su) d\su = 0 \,, 
\end{eqnarray}
is normalized to the maximum
value $\max\limits_{\su}\abs{\amp(\su)}=1$, and is periodic
$\amp(\su)=\amp(\su+\pi/\dimfreq)$ with the dimensionless frequency
\begin{eqnarray}
  \dimfreq \equiv  \frac{\freq}{4\shearrate_0}\,.
\end{eqnarray}
Thus, a frequency of $\dimfreq=1$ corresponds to a full rotation on the torus
in $\su$-direction.

We first show numerical results for the mean tumbling rate $\tumb$ as a
function of the driving frequency $\dimfreq$. These results were obtained by a
direct integration of the equations of motion (\ref{eq:skotheimscaled1}) and
(\ref{eq:skotheimscaled2}). Fig.~\ref{fig:resonancenum} shows the
characteristic dependence on the frequency for a harmonically oscillating
inverse shear rate. There is a smooth background, which is constant at low
frequencies and vanishes at a high cut-off frequency. A large number of
regularly ordered resonance peaks are superimposed.  A qualitative discussion
based on the results for time-constant shear rate can explain the general
features of this plot.

\subsection{Qualitative explanation}
\label{sec:qualitative-considerations}
\begin{figure}[t]
  \centering
  \includegraphics[width=0.99\linewidth]{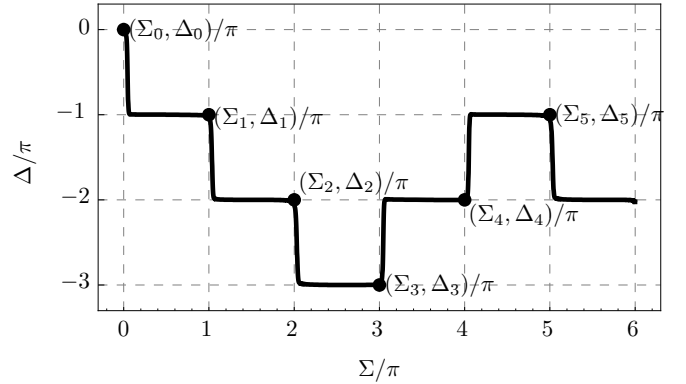}
  \caption{Sequence of junctions labeled by ($\su_k$, $\di_k$) for a given
    trajectory.}
  \label{fig:junctions}
\end{figure}

We start with some preliminary considerations, which will be confirmed
analytically afterwards. As shown in section~\ref{sec:matched-asymptotics} for
constant shear rates in the quasi-spherical limit, the equations of motion in
regions i) and iii) and the qualitative relaxation towards the stable line in
region iv) are independent of the shear rate. This behaviour remains unchanged
for a time-dependent shear rate or slope parameter $\slope(\su)$. Thus, the
relaxation towards the stable line and the motion on the stable line are
unaffected by the shear rate. As illustrated in Fig.~\ref{fig:junctions}, the
system therefore runs on the torus with monotonously increasing angle
$\su$. It moves close to the stable line into the junction region. Here it
turns to one vertical side depending on the value of $\slope$ and reaches a
neighbouring horizontal stable line. During the motion the junction region is
visited over and over again periodically in time $\dimtim$. We can label the
junction with angles ($\su_k$, $\di_k$), where
\begin{equation}
  \su_{k}\equiv k \pi
  \label{eq:defsuk}
\end{equation}
with integer index $k$ counts the number of visits (see
Fig.~\ref{fig:junctions}) and $\di_k$ is an integer multiple of $\pi$ counting
the difference of the number of tumbling and the number of swinging
motions. Starting with index $k=0$, the system reaches the junction at
consecutive angles $\su_0$, $\su_1$, $\su_2$, $\ldots$.

The only difference to the time-constant case of
section~\ref{sec:constant-shear-rate} happens in the junction region ii),
where the system leaves the stable line to follow the neutral line. Here, the
value of the instantenous shear rate determines for the overall
behaviour. Since the shear rate is now time-dependent, the slope parameter
$\slope$ can take different signs each time the system is in the junction
region and can even change signs several times within the junction region.

We first want to estimate the time the systems spends in the junction region
ii) and consider corresponding limit cases of the driving frequency
$\dimfreq$. Measuring time with respect to the nondimensional time $\dimtim$,
the speed of $\su$ is $\dimvisc$. The time the system needs to return to the
junction is of order $1/\dimvisc$. Since the junction region has linear
dimension of the order $\sqrt\dimvisc$, the time the system stays within the
junction region is of the order
$\sqrt\dimvisc/\dimvisc=1/\sqrt\dimvisc$. Thus, the fraction of time the
system is within the junction region is given by the order of $\sqrt\dimvisc$
and the corresponding frequency is of order $1/\sqrt\dimvisc$.

For high frequencies $\dimfreq\gg1/\sqrt\dimvisc$, the oscillation is too fast
for the system to respond. Therefore the shear rate behaves effectively as a
time-constant shear rate with mean slope $\slope_0$. For a negative mean slope
$\slope_0<0$, there is a pure swinging motion with vanishing mean tumbling
rate $\tumb=0$ (see Fig.~\ref{fig:resonancenum} b). Conversely, for a positive
mean slope $\slope_0>0$ there is a pure tumbling motion with mean tumbling
rate $\tumb=1$.

In the limit of low frequencies $\dimfreq\ll1/\sqrt\dimvisc$, the shear rate
in the junction region can be regarded constant. In other words, the junction
region is effectively just a point located at ($\su_k$, $\di_k$). Each time
the system is in the junction region labeled by $\su_k$, the sign of the slope
parameter $\slope(\su_k)$ determines whether the system performs a single
tumbling or a single swinging motion. The sign of $\slope(\su_k)$ depends on
the initial phase $\ph_0$, the frequency $\dimfreq$, and the index $k$. In
order to calculate the mean tumbling rate $\tumb$, the number of positive and
negative values of $\slope(\su_0)$, $\slope(\su_1)$, $\ldots$ have to be
counted. Therefore, the system can be mapped on a discrete model in the
low-frequency limit as shown in section \ref{sec:harm-oscill-shear}). This
discrete model reproduces both the constant background of the mean tumbling
rate and the superimposed resonance peaks which can both be seen in
Fig.~\ref{fig:resonancenum}.

For intermediate frequencies $\dimfreq\sim1/\sqrt\dimvisc$, it seems
reasonable that some time-averaged slope $\slope$ in the junction region
determines the motion of the capsule. This expectation is quantified in the
next section.

\subsection{Analytic solution in the junction region}
\label{sec:analytic-solution}
The qualitative arguments of the previous section are substantiated by a full
analytical investigation for a general time-dependent shear rate with
parameters $\dimvisc$ and $\diminvshear$ close to the special point. Since we
are in the quasi-spherical limit, we investigate the four regions analog to
the time-constant case (sec.~\ref{sec:matched-asymptotics}). The equations of
motion to lowest order in $\dimvisc$ remain unchanged except for the fact that
the slope parameter is now time-dependent $\slope=\slope(\su)$.  Since the
velocity in $\su$-direction in dimensionless units is $\dimvisc$, the system
visits the junction labeled by ($\su_{k}$, $\di_{k}$) at time $\dimtim=\pi
k/\dimvisc$.

As in the time-constant case (see Appendix~\ref{sec:matching}), there is no
dependence of the solutions on the slope parameter $\slope$ in regions i) and
iii), and no qualitative dependence on $\slope$ in region iv), where the
trajectory merely relaxes to the next stable line. Thus, after leaving the
junction region, the system moves fast towards one of the two neighbouring
stable lines, before moving slowly along the stable line and returning to the
junction region ii). Again, the junction region determines which stable line
is chosen next, i.e.~whether the capsule tumbles or swings. The corresponding
first order equation of motion (\ref{eq:ode2}) in region ii) close to
$\su_{k}$ can be integrated for a general time-dependent $\slope(\su)$ as is
shown in Appendix~\ref{sec:time-dependent-shear}. In the long-time limit
$\dimtim\gg1/\sqrt{\dimvisc}$ the matched solution (\ref{eq:switchtime}) in
the junction region becomes asymptotically
\begin{equation}
  \di(\su) - \di_k \approx 
  \sqrt{\frac{\pi\dimvisc}{2}}
  \exp{\left(\frac{2}{\dimvisc}(\su-\su_{k})^2\right)} \meanslope_{k}
\end{equation}
where the average slope $\meanslope_{k}$ corresponding to the $k$-th junction
($\su_k$, $\di_k$) is defined by
\begin{equation}
  \label{eq:defaverageslope}
  \meanslope_{k} \equiv \sqrt{\frac{2}{\pi\dimvisc}} 
  \int\limits_{-\infty}^{\infty} d\su
  \slope(\su)
  \exp{\left(-\frac{2}{\dimvisc}(\su-\su_{k})^2\right)}.
\end{equation}
This integral is a convolution of the time-dependent slope parameter
$\slope(\su)$ with a Gaussian shaped kernel of width $\sqrt\dimvisc$ centered
at $\su=\su_{k}$. Thus, for low frequencies $\dimfreq\ll 1/\sqrt\dimvisc$ the
kernel is effectively proportional to Dirac's $\delta$-function, while for
high frequencies $\dimfreq\gg 1/\sqrt\dimvisc$ the kernel smoothes out the
fast oscillations of $\slope(\su)$. These two limit cases will be discussed in
more detail in the next section.

The sign of the average slope $\meanslope_k$ determines whether the trajectory
of the system follows the neutral vertical line along the positive (for
$\meanslope_k>0$) or negative (for $\meanslope_k<0$) direction. Since the
solutions in regions iii) and iv) only describe the relaxation to the next
stable line, the asymptotic matching procedure then proceeds exactly as in the
time-constant case. For any given time-dependent shear rate $\shearrate(\tim)$
or equivalently $\slope(\su)$, the sequence of average slopes $\meanslope_{k}$
at $\su=\su_{k}$ can be calculated. The mean tumbling rate is then given by
\begin{equation}
  \label{meantumbling}
  \tumb = \lim_{N\rightarrow \infty}\frac{1}{N}
  \sum_{k=0}^{N-1} \Theta(\meanslope_{k})
\end{equation}
with the Heaviside step function $\Theta$. We now evaluate this expression for
a specific choice of $\slope(\su)$.

\subsection{Harmonically oscillating shear rate}
\label{sec:harm-oscill-shear}
Since $\slope(\su)$ is periodic with period $\pi/\dimfreq$, it can be
decomposed into a Fourier series consisting of an oscillation with the
fundamental frequency $2\dimfreq$ and the corresponding higher harmonics.  We
constrain $\slope(\su)$ to a pure harmonic oscillation in the following
section for simplicity. The results are easily generalised to the Fourier
series of an arbitrary periodic $\slope(\su)$ (see
Appendix~\ref{sec:fourier}).

\begin{figure*}[t]
  \centering
  \begin{minipage}{0.24\linewidth}
    a)\\\includegraphics[width=0.99\linewidth]{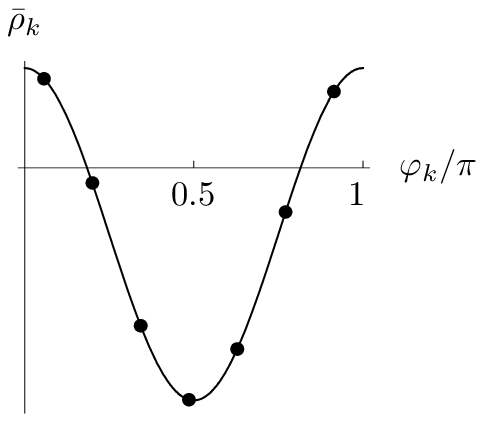}
  \end{minipage}
  \begin{minipage}{0.24\linewidth}
    b)\\\includegraphics[width=0.87\linewidth]{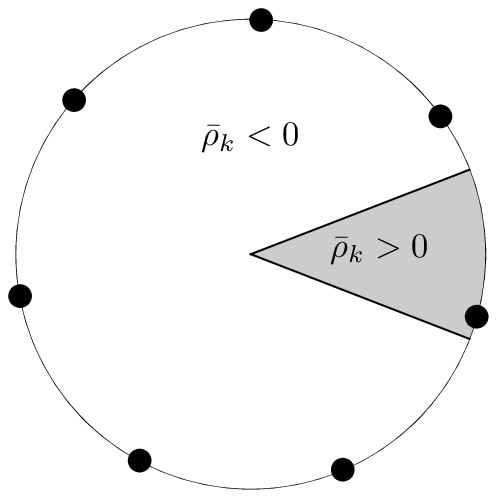}
  \end{minipage}
  \begin{minipage}{0.24\linewidth}
    c)\\\includegraphics[width=0.99\linewidth]{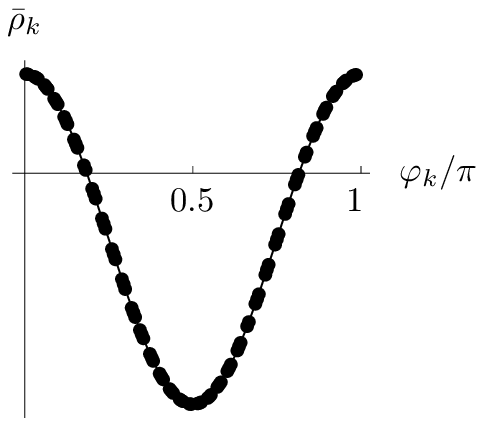}
  \end{minipage}
  \begin{minipage}{0.24\linewidth}
    d)\\\includegraphics[width=0.87\linewidth]{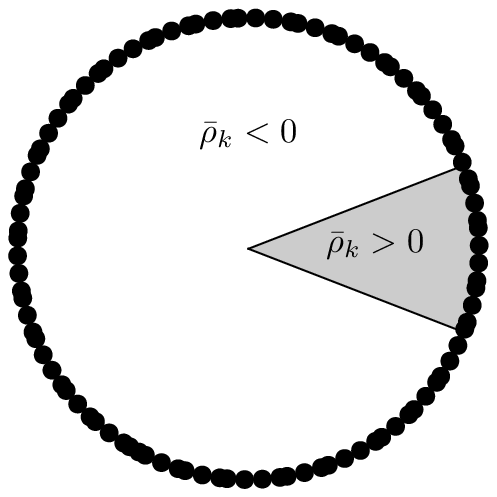}
  \end{minipage}
  \caption{a) and c) Mean slope $\meanslope_k$ as a function of the phase
    $\ph_k$ for negative $\slope_0<0$ and rational (a) or irrational (c)
    frequency. b) and d) Corresponding sequences of phases and intervals in
    which $\meanslope_k(\ph_k)$ is positive and negative on the circle $S^1$.}
  \label{fig:circles}
\end{figure*}

For a purely harmonically modulated slope
\begin{equation}
  \slope(\su) 
  \equiv \slope_0 + \slope_1 \cos{(2(\dimfreq \su + \ph_0))}
  \label{eq:osci2}
\end{equation}
with mean value $\slope_0$, amplitude $\slope_1 > 0$, frequency $\dimfreq$ and
initial phase $\ph_0$, the sequence of mean slopes $\meanslope_{k}$ can be
evaluated analytically
\begin{equation}
  \meanslope_{k} = \slope_0 + 
  \exp{\left(-\frac{\dimvisc\dimfreq^2}{2}\right)}  \slope_1 \cos(2\ph_k)\,,
  \label{eq:meanslope}
\end{equation}
where we have defined the sequence of equidistant phases
\begin{equation}
  \ph_{k}\equiv \ph_{0}  + k \pi \dimfreq \,.
\end{equation}
We now discuss the mean tumbling rate
\begin{equation}
  \tumb = \lim_{N\rightarrow\infty} \sum_{k=0}^{N-1} 
  \Theta\left(\slope_0 + \exp{\left(-\dimvisc\dimfreq^2/2\right)} 
    \slope_1 \cos(2\ph_k)\right)    
\end{equation}
as a function of $\slope_{0}$, $\slope_{1}$, $\dimfreq$, and $\ph_{0}$.

For small modulation amplitudes $\exp\left(-\dimvisc\dimfreq^2/2\right)
\slope_{1} < |\slope_{0}|$, the sign of $\slope_{0}$ equals the sign of
$\meanslope_{k}$ for all $k$. This condition holds for all frequencies if
$\slope_{1} < |\slope_{0}|$. For $\slope_{1} > |\slope_{0}|$, we can define a
threshold frequency
\begin{equation}
   \dimfreq_c 
   \equiv {\dimvisc}^{-1/2}
   \sqrt{{2\ln{\left(\frac{\slope_1}{\abs{\slope_0}}\right)}}},
   \label{eq:thresholdfreq}
\end{equation}
beyond which ($\dimfreq \ge \dimfreq_{c}$) the phase behaviour is given by the
sign of $\slope_{0}$ alone. The system tumbles ($\tumb=1$) for $\slope_{0}>0$
and swings ($\tumb=0$) for $\slope_{0}<0$.

For large modulation amplitudes $\exp\left(-\dimvisc\dimfreq^2/2\right)
\slope_{1} > |\slope_{0}|$, the sign of $\meanslope_{k}$ depends on the value
of the phase $\ph_{k}$ modulo $\pi$. For phases in the region $|\ph_{k}| \leq
\Delta \ph/2$ with
\begin{eqnarray}
  \Delta \ph &\equiv& 
  \arccos\left(-\frac{\slope_0}{\slope_1}
    \exp\left(\frac{\dimvisc\dimfreq^2}{2}\right)\right)\,,
  \label{eq:deltaph}
\end{eqnarray}
the average slopes $\meanslope_k$ are positive and the system performs
tumbling motions and swinging motions otherwise. This behaviour is visualised
in Fig.~\ref{fig:circles}, where the phases $\ph_{k}$ modulo $\pi$ are
interpreted as points $\exp(2i\ph_{k})$ on the circle $S^{1}$ (modulo $\pi$).

The circle $S^1$ consists of an arc with angle $\Delta\ph$ corresponding to
tumbling and a complementary arc with angle $\pi - \Delta\ph$ corresponding to
swinging. Each phase $\ph_k$ at the junction labeled by ($\su_k$, $\di_k$) is
either an element of the tumbling or an element of the swinging arc. By
counting the fraction of phases within each arc the mean tumbling rate can be
calculated explicitly:
\begin{itemize}
\item For an irrational frequency $\dimfreq$ the values of the phases
  $\ph_k=\ph_0+k\pi\dimfreq$ lie densely on the circle $S^1$. The fraction of
  number of swinging to tumbling motions in the long time limit is then given
  by the ratio of the length $\Delta\ph$ and $\pi-\Delta\ph$ of the two
  intervals, leading to a mean tumbling rate of
  \begin{equation}
    \tumb = \frac{\Delta \ph}{\pi} = 
    \frac{\arccos{
        \left(-\slope_0\exp(\dimvisc\dimfreq^{2}/2)/\slope_1\right)}}{\pi}
    \,.
    \label{eq:background}
  \end{equation}
  In the low frequency limit this becomes a constant $\tumb \approx
  \arccos(-\slope_{0}/\slope_{1})/\pi$.

\item For a rational frequency $\dimfreq = p/q$ with integer and coprime
  numbers $p$ and $q$ the phases $\ph_k = \ph_0 + k \pi \dimfreq = \ph_0 +
  {kp}\pi/{q}$ lie on $q$ equidistant phases $\ph_0+ j \pi /q $ (with
  $j=0,\ldots,q-1$) on the circle $S^1$. In the long time limit, all of these
  $q$ angles are visited equal amounts of times. The ratio of number of
  swinging to tumbling motions is given by the ratio of number phases $\ph_k$
  in the two intervals of length $\Delta\ph$ and $\pi - \Delta \ph$. For high
  values of $q$, approximately $q\Delta\ph$ phases lie within the interval
  $\Delta\ph$. Then the mean tumbling rate $\tumb$ is approximately given by
  the expression~(\ref{eq:background}) valid for irrational frequencies. For
  low denominators $q$, the number of phases within the two intervals
  additonally depends upon the initial phase $\ph_0$. An integer frequency
  $\dimfreq=p$ for instance gives a tumbling rate of either $\tumb = 1$ or
  $\tumb = 0$ depending only on the initial condition. For a general rational
  frequency, counting the number of phases $|\ph_{k}|\le \Delta\ph$ in the
  tumbling sector gives
  \begin{equation}
    \tumb = \frac{1}{q} \floor{\frac{q}{\pi} \left(\frac{\Delta
          \ph}{2} - \ph_{0}\right)} 
    + \frac{1}{q} \ceil{\frac{q}{\pi} \left(\frac{\Delta
          \ph}{2} + \ph_{0}\right)},\label{eq:resonance}
  \end{equation}
  with $\Delta \ph$ given by eqn.~(\ref{eq:deltaph}). Here, $\floor{~}$ and
  $\ceil{~}$ are the floor and ceiling functions, which denote the closest
  integer smaller or larger than the argument, respectively.
\end{itemize}
Plotting the mean tumbling rate $\tumb$ over small frequencies $\dimfreq$ in
Fig.~\ref{fig:resonancenum}, we can identify a smooth irrational background
(\ref{eq:background}) superimposed by rational peaks
(\ref{eq:resonance}). Their amplitudes depends on the denominator $q$ and the
initial phase $\ph_0$. For integer resonance frequencies, the peaks go either
to $0$ or $1$.
\begin{figure}[t]
  \centering
  \begin{minipage}{0.99\linewidth}
    a)
    \begin{center}
      \includegraphics[width=0.84\linewidth]{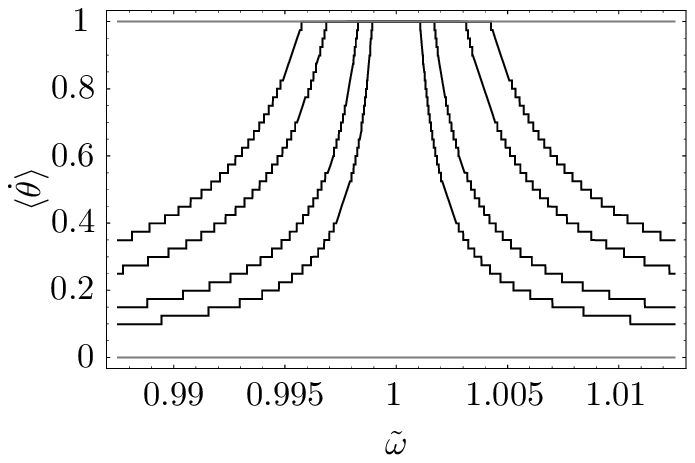}
    \end{center}
    b)
    \begin{center}
      \includegraphics[width=0.84\linewidth]{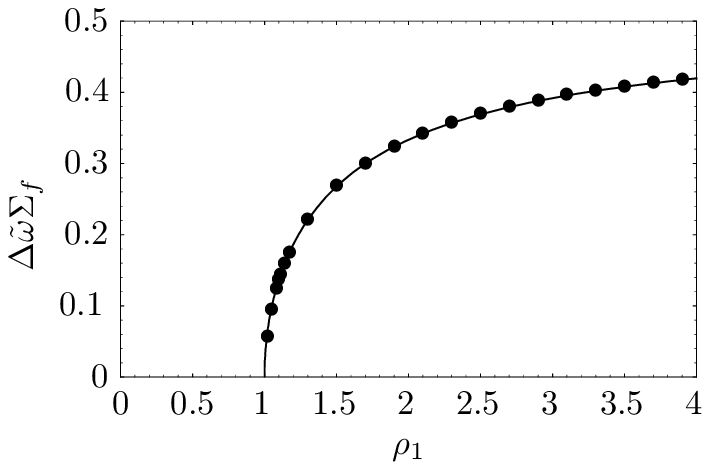}
    \end{center}
  \end{minipage}
  \caption{a) Mean tumbling rate $\tumb$ at the rational peak corresponding to
    $\dimfreq=1$ for different slope amplitudes $\slope_1=2.$, $1.4$, $1.1$,
    $1.04$. The higher the amplitude $\slope_1$, the smaller the width
    $\Delta\dimfreq$. The constant parameters are $\ecc = 0.001 \pi$,
    $\dimvisc = 0.01$, $\slope_0 = -1$, a finite running time $\su_f=40\pi$,
    and the initial phase $\ph_0$ within the tumbling arc.  b) Numerical
    (dots, from a) with $\tumb>0.999$) and analytical (solid line,
    eqn.~(\ref{eq:width})) results of the peak widths $\Delta\dimfreq$ for
    different amplitudes $\slope_1$ (remaining parameters as in a).}
  \label{fig:width}
\end{figure}

Mathematically speaking, the peaks at rational frequencies are infinitesimally
narrow. Since experiments and numerical methods run only for a finite time,
the peaks are broadened to a finite width, which can be estimated in the
following for the most dominant peaks with nearly integer frequency
\begin{equation}
  \dimfreq = p + \delta \dimfreq.  
\end{equation}
Let us assume that $\ph_{0}=0$, so that the system tumbles with $\tumb = 1$
(for $\dimfreq=p$). For exactly integer modulation frequency $\delta
\dimfreq=0$, the phase stays constant $\ph_{k}=0$ for all $k$.  For a
non-integer frequency $\delta\dimfreq\neq0$ the $k$-th phase modulo $\pi$ is
$\ph_{k}= k\pi\delta\dimfreq = \delta\dimfreq \su_{k}$. If this phase does not
change by more than by the width of the tumbling arc $\Delta \ph$ during the
simulation time $\su_{f}$, the system stays in the tumbling regime. Therefore
we can estimate the width of the integer resonance peaks as
\begin{equation}
  \Delta \dimfreq 
  \simeq \frac{\Delta \ph}{\su_{f}} 
  = \frac{1}{\su_{f}}\arccos\left(-\frac{\slope_0}{\slope_1}
    \exp\left(\frac{\dimvisc\dimfreq^2}{2}\right)\right) \,.  
\label{eq:width}
\end{equation}
This result is confirmed in Fig.~\ref{fig:width}, where numerically evaluated
peak widths are compared to expression~(\ref{eq:width}) for different
modulation amplitudes $\slope_{1}$. For the numerical data, the peak width was
defined to be the frequency interval in which the mean tumbling rate during
the simulation time $\tumb$ was larger than $0.999$. The agreement is
excellent.

\subsection{Dynamic phase diagram}
\begin{figure}[t]
  \centering
  \includegraphics[width=0.94\linewidth]{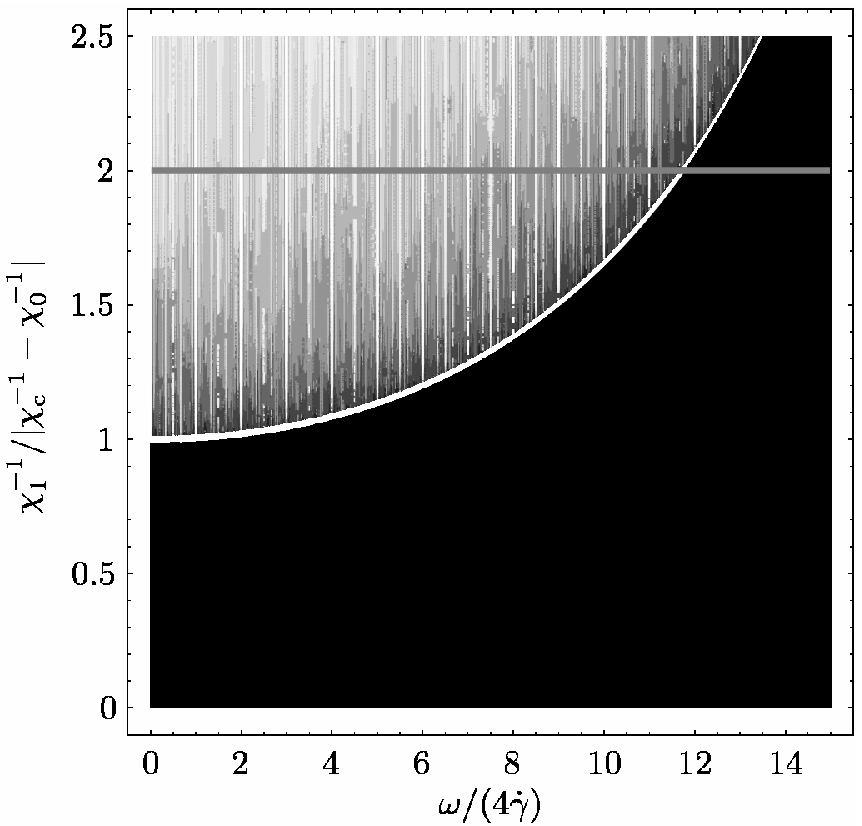}
  \caption{Dynamic phase diagram visualized as a grey scale plot of the mean
    tumbling rate $\tumb$ as a function of oscillation amplitude relative to
    the distance of the mean shear rate to the phase boundary
    $\dimshear_{1}^{-1}/\abs{\diminvshear_{\text{c}}-\diminvshear_{0}}$, and
    of the normalised modulation frequency $\freq/(4\shearrate_{0})$. Black
    colour indicates pure swinging, $\tumb=0$. Tumbling only occurs above a
    threshold in the oscillation amplitude given by
    eqn.~(\ref{eq:oscthreshold}). Data shown in Fig.~\ref{fig:resonancenum}
    correspond to a cut in parameter space as indicated by the grey
    horizontal line.}
  \label{fig:dynphase}
\end{figure}

We summarise our findings in a dynamic phase diagram. While the specific shape
of the phase diagram depends on the particular functional time-dependence of
the shear flow, the general features hold for any oscillating time-dependent
flow with mean inverse shear rate $\diminvshear_{0}$ and oscillation amplitude
$\diminvshear_{1}$, compare eqn.~(\ref{eq:osci1}). From our study of
time-constant shear flow we know the location of the phase boundary
$\diminvshear_{\text{c}}=1-\dimvisc/2$. In Fig.~\ref{fig:dynphase}, a grey
scale plot of the mean tumbling rate $\tumb$ is shown as a function of the
oscillation amplitude $\diminvshear_{1}$ relative to the distance of the mean
shear rate to the phase boundary,
$\diminvshear_{1}/|\diminvshear_{\text{c}}-\diminvshear_{0}|$, versus the
oscillation frequency $\dimfreq=\freq/4\shearrate_{0}$ measured im units of
the mean shear rate. The colour level indicates the mean tumbling rate as
defined in eqn.~(\ref{eq:tumb}) for finite simulation times, black colour
indicating a swinging motion $\tumb=0$. In Fig.~\ref{fig:dynphase}, a mean
shear rate in the swinging regime (for constant flow) was chosen,
$\diminvshear_{0}<\diminvshear_{\text{c}}$. One can see that for oscillation
amplitudes below the distance to the phase boundary,
$\diminvshear_{1}<\diminvshear_{c}-\diminvshear_{0}$, the capsule never
tumbles. In order to induce tumbling motion, the instantaneous shear rate has
to cross the phase boundary. For higher modulation frequencies, the
oscillation amplitude threshold for tumbling is even higher and given by
\begin{equation}
  \label{eq:oscthreshold}
  \diminvshear_{1,\text{c}} \equiv 
  \abs{\diminvshear_{\text{c}}-\diminvshear_{0}}
  \exp\left(\frac{\dimvisc\freq^{2}}{32\shearrate^{2}}\right).
\end{equation}
Above the oscillation amplitude threshold the mean tumbling rate grows
continuously on the irrational background with increasing amplitude. At the
resonance frequencies the mean tumbling rate reaches values given by
eqn.~(\ref{eq:resonance}).

\section{Conclusions}
\label{sec:conclusions}
We have investigated the motion of microcapsules in time dependent shear flow
in a reduced model. The equations of motions were studied analytically in the
quasi-spherical limit for constant viscosity contrast. We have identified the
stable and unstable fixed points at lowest order in the deformation, which
lead to swinging and tumbling limit cycles at first order depending on the
shear rate. Close to the dynamic phase boundary, the expansion of the
equations of motion was carried out to first order in the deformation. Their
analytic solution was facilitated by solving the trajectories in different
regions in phase space separately: One region close to the stable fixed points
with comparatively slow dynamics, one region close to the unstable fixed
points with a comparatively fast motion, and a junction close to the neutral
fixed point, which also acts on a fast time scale. The direction of the
dynamics in this latter region is determined by the value of the shear
rate. The analytic inner solutions for the trajectories were then joined
together using the method of matched asymptotic expansion.

As a central result of this study, we now fully understand the dynamic phase
behaviour of quasi-spherical capsules in time-independent shear flow and have
determined the phase boundary between swinging and tumbling as a function of
shear rate and viscosity as $\dimshear^{-1}=1-\dimvisc/2$. In physical
parameters the phase boundary reads 
\begin{eqnarray}
  \shearrate^{-1} = \frac{\vol\viscout\sm}{2
    \elast}\left[10-\sm\left(3+2\frac{\viscin}{\viscout}\right)\right] \,.
\end{eqnarray}

We then generalised our result to time-dependent shear rates. The equations of
motion were integrated analytically in the different dynamic regions. Again,
we found that the motion for one period is determined by a weighted
time-average of the shear rate during the time when the system is close to the
junction. We have thus mapped the continuous capsule dynamics to a discrete
model valid for all time-dependent flows. Our general results were then
applied to harmonically modulated shear rates around a finite mean value,
where the dynamic phase diagram was constructed explicitly. As the dynamics is
determined by the (time-averaged) value of the shear rate at specific times
only, the system shows a pronounced resonance behaviour: By choosing suitable
modulation frequencies, it is possible to induce tumbling motion for capsules,
which would otherwise swing at constant mean shear rate. The width of the
resonance peaks for finite simulation time can also be expressed
analytically. For not resonant frequencies, the mean tumbling rate is
determined by an analytic expression, which vanishes beyond an upper
modulation frequency threshold. The agreement of numerical solutions of the
original equations of motion with the theoretical results is excellent.

In summary, we have reached a fairly complete analytical understanding of the
motion of a quasi-spherical capsule in time-dependent shear flow. In the
course of the study, we have also obtained analytical expressions for capsules
in constant shear flow whose equations of motion previously have been studied
merely numerically.

\begin{acknowledgement}
  Financial support of the DFG with in the priority programme SPP 1164 ``Nano-
  and Microfluidics'' is gratefully acknowledged.
\end{acknowledgement}

\begin{appendix}

\numberwithin{equation}{section}
\renewcommand\theequation{\Alph{section}.\arabic{equation}}

\section{Keller-Skalak quantities}
\label{sec:keller-skalak}
The geometrical quantities $\f_i$ occuring in equations (\ref{eq:skotheim1})
and (\ref{eq:skotheim2}) depend upon the semi-axes $\axis_1$, $\axis_2$,
$\axis_3$ and are defined by the following relations:
\begin{eqnarray}
  r_2 &\equiv& \frac{\axis_2}{\axis_1} \,, ~~
  r_3 \equiv \frac{\axis_3}{\axis_1} \,, ~~
  \alpha_i \equiv 
  \frac{\axis_i}{\left(\axis_1\axis_2\axis_3\right)^{1/3}} \,,\\
  \tilde \Delta &\equiv& 
  \sqrt{(\alpha_1^2+s)(\alpha_2^2+s)(\alpha_3^2+s)} \,,\\
  g'_3 &\equiv& 
  \int\limits_0^\infty\frac{ds}{(\alpha_1^2+s)(\alpha_2^2+s)\tilde\Delta} \,,\\
  z_1 &\equiv& \frac{1}{2r_2} - \frac{r_2}{2} \,, ~~
  z_2 \equiv g'{}_3(\alpha_1^2+\alpha_2^2) \,,\\
  \f_1 &\equiv& \left(r_2-\frac{1}{r_2}\right)^2 \,, ~~
  \f_2 \equiv 4z_1^2\left(1-\frac{2}{z_2}\right) \,,\\
  \f_3 &\equiv& -4\frac{z_1}{z_2} \,.
\end{eqnarray}
In the axisymmetric cases $\axis_2=\axis_3$ or $\axis_1=\axis_3$ and in the
quasi-spherical case $\abs{1-\axis_{2,3}/\axis_1}\sm\ll1$, the integral $g'_3$
can be computed explicitly.

\section{Notation for asymptotic limits}
\label{sec:notation}
We use the following conventions, where $f(x)$ and $g(x)$ are real functions
of the real variable $x$, $c$ is a real constant and $y$ and $z$ are real
numbers:
\begin{eqnarray}
  &&
  f(x) \sim g(x) 
  ~~~ \Leftrightarrow ~~~
  \lim\limits_{x\to 0} \frac{f(x)}{g(x)} = c \,,\\
  &&
  f(x) \approx g(x) 
  ~~~ \Leftrightarrow ~~~
  \lim\limits_{x\to 0} \frac{f(x)}{g(x)} = 1 \,,\\
  &&
  y \simeq z 
  ~~~ \Leftrightarrow ~~~
  y \text{~and~} z \text{~numerically~equal}\,.
\end{eqnarray}

\section{Asymptotic matching for time-constant shear rates}
\label{sec:matching}
Here, we describe our analytical procedure for obtaining the asymptotic
trajectory of the limit cycle in the quasi-spherical case
$\dimvisc\sim\sm\ll1$ when the system is close to the special point
$\dimvisc=0$ and $\diminvshear=1$. For each of the four regions of
Fig.~\ref{fig:fixedpoints2}, the quasi-spherical equation of motion
(\ref{eq:central-asymptotic}) is solved asymptotically, compare
Fig.~\ref{fig:asymptotic}.
\begin{itemize}
\item[i)] $\su\sim1$, $\di\sim\sqrt\dimvisc$: Expansion gives the equation of
  motion
  \begin{eqnarray}
    \frac{d \di}{d \su} 
    = \frac 4 \dimvisc \di\sin\su
    \label{eq:ode1}
  \end{eqnarray}
  which is independent of the shear rate, i.e.~independent of the slope
  $\slope$. Its solution is given by
  \begin{eqnarray}
    \di(\su) = 
    \di_1 \exp{\left(-\frac4\dimvisc(\cos{\su}-\cos{\su_1})\right)} \,,
    \label{eq:sol-reg-1}
  \end{eqnarray}
  where $(\su_1,\di_1)$ is an arbitrary point on the trajectory. This
  solution will be used to match with the solution of region ii). As can be
  seen in iv) by closing the trajectory on the torus, the limit cycle in
  region i) is simply given by the original stable line
  \begin{eqnarray}
    \di(\su)=0 \,.
    \label{eq:region1}
  \end{eqnarray}
  Thus, to leading order in $\dimvisc\sim\sm$, the system runs on the stable
  line $\di(\su)=0$ with $-\pi<\su<0$. Even when the system starts off the
  limit cycle, the stable character of $\di=0$ and $-\pi<\su<0$ leads to a
  fast relaxation towards the stable line, while the angle $\su$ changes
  slowly due to $\dimvisc\ll1$. Therefore, the matching with region ii) will
  not depend on the initial point $(\su_1,\di_1)$ and the general solution
  (\ref{eq:sol-reg-1}) is independent of the slope parameter $\slope$.
\item[ii)] $\su\sim\sqrt\dimvisc$, $\di\sim\sqrt\dimvisc$:
  Here, stable and neutral line meet, and the expansion gives
  \begin{eqnarray}
    \frac{d \di}{d \su} 
    = \frac 4 \dimvisc \di\su + \slope \,.
    \label{eq:ode2}
  \end{eqnarray}
  In this region, the vector field corresponding to $\dimvisc=0$ is small
  enough for the finite but small value of $\dimvisc$ to have a significant
  influence on the the vector field and thus on the motion. Here, the exact
  value of the the shear rate or the slope parameter $\slope$ is critical as
  can be seen by the solution
  \begin{eqnarray}
    \di(\su) &=& 
    \sqrt{\frac {\pi\dimvisc} 8 } \slope
    \left(1+\erf{(\sqrt{\frac 2 \dimvisc} \su)}\right) \nonumber\\
    &&\times \exp{\left(\frac 2 \dimvisc\su^2\right)} \,,
    \label{eq:region2}
  \end{eqnarray}
  which has been matched with the general solution (\ref{eq:sol-reg-1}) of
  region i) and which is independent of the initial point. Here, we use the
  error function
  \begin{eqnarray}
    \erf{x} \equiv \frac{2}{\sqrt\pi}\int\limits_0^x ds \exp{(-s^2)} \,.
  \end{eqnarray}
  Thus, the sign of $\slope$ determines the sign of $\di(\su)$. For negative
  $\slope<0$ the neutral line with $\di<0$ is choosen which leads to a
  swinging motion. Conversely, for a positive $\slope>0$ the neutral line with
  $\di>0$ is choosen, corresponding to a tumbling motion (see region iii)).
  Region ii) with $\su\sim\di\sim\sqrt{\dimvisc}$ acts as a junction which only
  depends on the sign of the slope parameter $\slope$.
\item[iii)] In the region $\su\sim\sqrt\dimvisc$ and $\di\sim1$ close to the
  neutral line, the expansion gives a slope-independent equation of motion
  \begin{eqnarray}
    \frac{d \di}{d \su} 
    = \frac 4 \dimvisc \su\sin\di \,,
    \label{eq:ode3}
  \end{eqnarray}
  whose matched solution
  \begin{eqnarray}
    \label{eq:region3}
    \di(\su) &=& 
    2\arctan \left(
      \tan{(\sqrt{\frac {\pi\dimvisc} 8} \slope)} 
    \right.
    \\
    &&\left.\times
    \exp{\left(\frac 2 \dimvisc\su^2\right)}
    \right) 
    \nonumber
  \end{eqnarray}
  depends on the slope parameter $\slope$. This solution describes the
  relaxation towards the next stable line which has been chosen in region ii).
\item[iv)] In the region $\su\sim\sqrt\dimvisc$ and
  $\di\pm\pi\sim\sqrt\dimvisc$, where neutral and unstable lines meet again,
  the expansion gives
  \begin{eqnarray}
    \frac{d \di}{d \su} 
    = - \frac 4 \dimvisc (\di\pm\pi)\su + 2 - \slope
    \label{eq:ode4}
  \end{eqnarray}
  with matched solution
  \begin{eqnarray}
    &&\di(\su) = 
    \mp\pi +  
    2\exp{\left(-\frac 2 \dimvisc \su^2\right)} \times 
    \label{eq:region4}
    \\
    &&\left(
      \pm \pi - \cot(\sqrt{\frac {\pi\dimvisc} 8} \slope)
      + \frac{\sqrt{2\pi\dimvisc}}8 
      (2-\slope) \erfi{(\sqrt{\frac 2 \dimvisc} \su)}
    \right)
    \,, \nonumber
  \end{eqnarray}
  where we used the imaginary error function
  \begin{eqnarray}
    \erfi{x} \equiv \frac{\erf{(ix)}}{i} \,.
  \end{eqnarray}
  The upper and lower signs depend on which neutral line was chosen in region
  ii). Although both the equation of motion and the solution depend on
  $\slope$, the system relaxes toward the stable line for all values of
  $\slope$. It then starts over again in region i) close to the stable
  line. By matching with the general solution of region i) the trajectory
  closes and the solution (\ref{eq:region1}) is obtained.
\end{itemize}

\section{Asymptotic matching for time-dependent shear rates}
\label{sec:time-dependent-shear}
The first order equation of motion (\ref{eq:ode2}) in region ii)
\begin{eqnarray}
  \frac{d \di}{d \su} 
  = \frac 4 \dimvisc \di\su + \slope(\su)
  \label{eq:ode2time}
\end{eqnarray}
can be integrated for a general time-dependent $\slope(\su)$. In order to
match asymptotically with the solution $\di(\su)=0$ (\ref{eq:region1}) of
region i), the limit $\dimtim\to-\infty$ or equivalently $\su\to-\infty$ has
to be taken. The matched solution in the junction region then is given by
\begin{eqnarray}
  \di(\su) &=& 
  \exp{\left(\frac2\dimvisc\su^2\right)} \times
  \label{eq:switchtime} \\
  &&\int\limits_{-\infty}^\su d\su' \slope(\su')
  \exp{\left(-\frac2\dimvisc\su'^2\right)} \,.
  \nonumber
\end{eqnarray}

For a harmonically changing slope parameter (\ref{eq:osci2}) with mean value
$\slope_0$, amplitude $\slope_1$, frequency $\dimfreq$ and initial phase
$\ph_0$ the matched solution (\ref{eq:switchtime}) can be integrated
\begin{eqnarray}
  \di(\su) &=& -\frac{1}{2}\sqrt{\frac{\pi\dimvisc}{2}}
  \exp{\left(\frac{2\su^2}{\dimvisc}\right)} 
  \label{eq:solution}
  \\
  && \times
  \left[
    \slope_0 \erfc{\left(\sqrt{\frac{2}{\dimvisc}}\su\right)} 
    +\slope_1 \exp{\left(-\frac{\dimvisc\dimfreq^2}{2}\right)}\right.
    \nonumber\\
  &&  
  \left(
  \cos 2\ph_0\re{\left(
    \erfc{\left(\frac{4\su+i\dimvisc\dimfreq}{\sqrt{2\dimvisc}}\right)}
    \right)}
  \right. \nonumber\\
  && \left.\left.
  +\sin 2\ph_0\im{\left(
    \erfc{\left(\frac{4\su+i\dimvisc\dimfreq}{\sqrt{2\dimvisc}}\right)}
    \right)}
  \right)
  \right] \nonumber \,,
\end{eqnarray}
where $\erfc{z}\equiv 1-\erf{z}$ is the complementary error function, $\re$
and $\im$ denote real and imaginary parts, respectively.

\section{Fourier series}
\label{sec:fourier}
The oscillating part $\amp(\su)$ of any general period slope $\slope(\su)$
(see eqn.~(\ref{eq:osci1})) can be uniquely decomposed into a Fourier series
\begin{equation}
  \amp(\su) = \sum\limits_{j=1}^\infty \amp_j \cos(2(j \dimfreq \su + \aph_j))
   \,.
\end{equation}
In this case, the sequence of mean slopes $\meanslope_k$ (see
eqn.~(\ref{eq:defaverageslope})) is explicitly given by the Fourier series
\begin{equation}
  \meanslope_{k} = \slope_0 + \slope_1
  \sum\limits_{j=1}^\infty \amp_j \exp{\left(-\frac{\dimvisc j^2
        \dimfreq^2}{2}\right)} \cos(2(j \dimfreq k\pi + \aph_j)) \,.
\end{equation}
For $b_1=1$ and $b_j=0$ for all $j>1$, we recover eqn.~(\ref{eq:meanslope}).
The amplitude $\amp_j$ is damped by the factor $\exp{(-\dimvisc j^2
  \dimfreq^2/2)}$.

\end{appendix}

\end{document}